\documentclass{aa}
  \usepackage{graphicx}
  \begin{document}
 \title{Kinematics and flux evolution of superluminal components
     in QSO {\bf{B}}1308+326}
 \author{S.J.~Qian\inst{1}}
 \institute{National Astronomical Observatories, Chinese
    Academy of Sciences, Beijing 100012, China}
 \date{Compiled by using A\&A-latex}
 \abstract{Search for Doppler-boosting effect in flux evolution
   of superluminal components in blazars has been an 
  important subject, which can help to clarify their kinematic and emission
    properties.}{The kinematics and flux evolution observed at 15\,GHz for
   the three superluminal components (knot-c, -i and -k) in QSO B1308+326 
   (z=0.997) were investigated in detail.}{It is shown that the precessing jet
   nozzle model previously proposed by Qian et al. (1991, 2014, 2017, 2022a,
   2022b) can be 
   used to fully simulate their kinematics on pc-scales with a nozzle 
   precession period of 16.9\,yr.}{With the acceleration/deceleration
    in their motion found in the model-simulation of their kinematics 
    we can derive their bulk Lorentz factor and Doppler factor as function 
   of time and predict their Doppler-boosting effect.} 
   {Interstingly, the flux evolution of the three superluminal components 
   can be well intrepreted in terms of their  Doppler-boosting effect. 
   The full explanation of both their kinematic behavior and flux evolution 
   validates our precessing nozzle
   model and confirms that superluminal components are physical entities moving
   relativistically toward us at small viewing angles.} 
  \maketitle
    \section{Introduction}
     B1308+326 is a high redshift quasar (z=0.997). In historical records
    it was observed as being optically variable with a long-term variability
     amplitude of $M_{B}\sim$5.6\,mag and highly polarization and was classified
     as one of the most variable BL Lac objects (Angel \& Stockman \cite{An80}).
     It is a $\gamma$-ray source detected by the  
    "Fermi Gamma-ray Observatory" (Ackermann et al. \cite{Ack11},
     Acero et al. \cite{Ac15}). So B1308+326 radiates across the entire 
     electromagnetic spectrum from radio-mm-NIR-optical through X-ray to 
     $\gamma$-ray bands. Very strong variability has been observed in all
     these wavebands  with various timescales (hours/days to years).\\
     B1308+326 is a low-synchrotron-peaked high polarization 
     quasar, showing a radio core-jet structure on pc-scales with superluminal
    components ejected from its core. 
    In the previous works (Qian et al. \cite{Qi17}, Britzen et al. \cite{Br17})
   the kinematic behaviors of the five superlumianl components (knot-c, -h, -i,
   -j and -k) observed  at 15\,GHz in B1308+326 were 
   well model-simulated in terms of our precessing nozzle scenario with  
   a precession period of 16.9$\pm$0.85 years, revealing their motion along the
   precessing common trajectory in the innermost jet regions corresponding
    to their precession phases.\\
     The interesting results (including the modeled distribution of
   the precessing common trajectories, the model-fits of the observed 
   trajectories, the periodic swing of their ejection position angle and  the 
   relation between the position angle and their initial viewing
    angle) are re-plotted in Figures 1 and 2. Although these results
    are "very nice\,!!" as an anonymous referee commented, they 
    are not consummate, because the bulk Lorentz factor and Doppler 
   factor derived for the superluminal knots in our model-simulations
    were not further investigated to find out
   the association of their flux evolution with the Doppler-boosting
   effect. Therefore, one would ask the questions: 
   What about the flux-density evolution of the superluminal knots ?
   Would the evolving line-of-sight of the feature  trajectories and 
   their bulk Lorentz factor produce measurable changes in the 
   flux densities? Would  the used precession model be able to predict 
   (at least part of)  the light curves of the different componnets ? \\
   In this paper we shall  discuss the flux evolution combined with  
   the model-fitting of the kinematic behavior for three superluminal 
   components (knot-c, -i and -k), demonstrating that their Doppler-boosting 
   effect plays an important role for understanding their kinematic, dynamic 
    and emission properties. That is, the inclusion of flux evolution can
   help to accurately interpret their VLBI-kinematics, correctly deriving
   their bulk Lorentz factor and Doppler factor as function of time in the 
   model-simulations and showing that  their flux evolution can be well 
   explained in terms of the Doppler-boosting effect. 
     \begin{figure*}
     \centering
     \includegraphics[width=5.6cm,angle=-90]{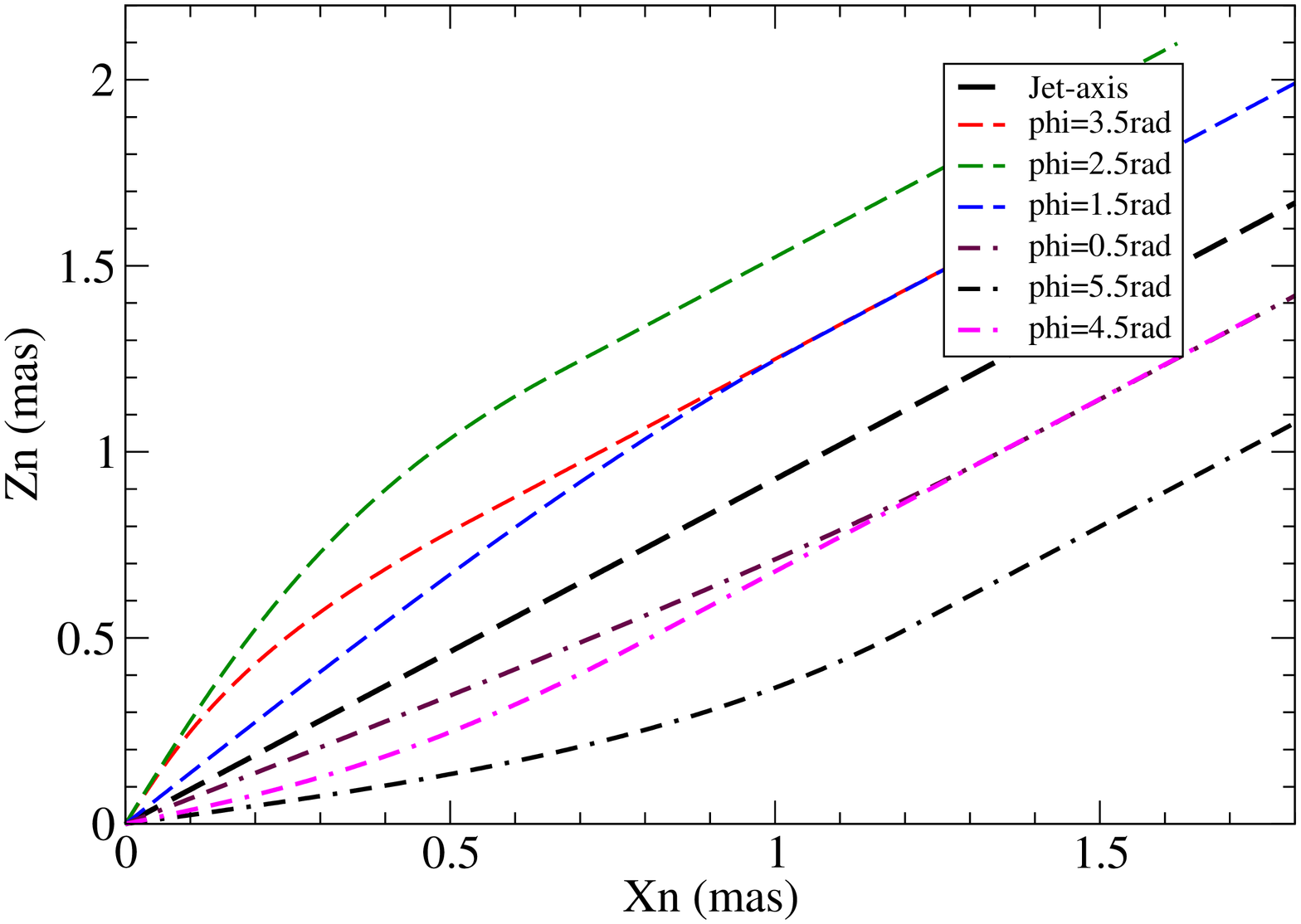}
     \includegraphics[width=5.6cm,angle=-90]{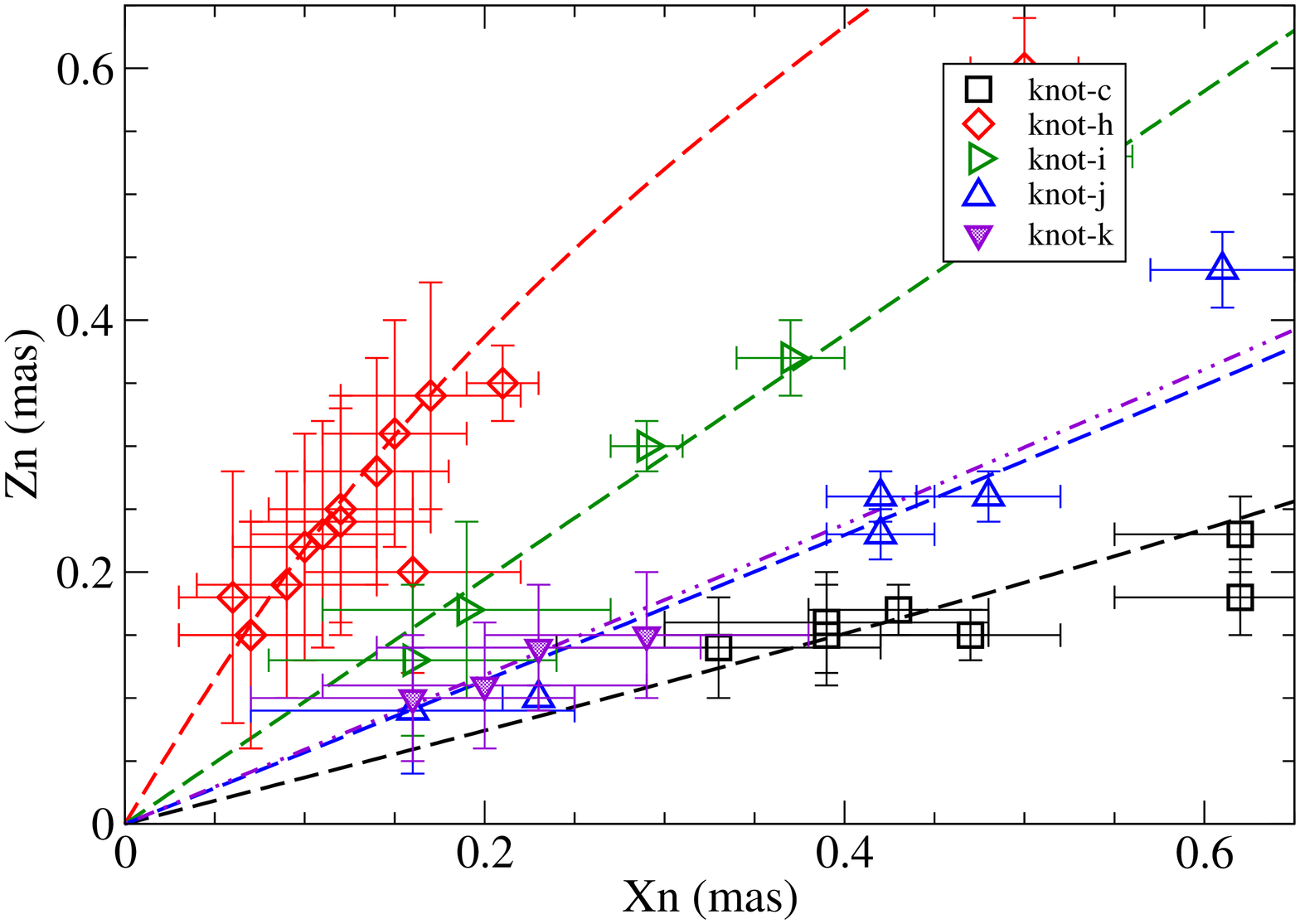}
     \caption{Left panel: the distribution of precessing common trajectories
     modeled for superluminal components in B1308+326 at precession phases
     $\phi$=0.5, 1.5, 2.5, 3.5, 4.5 and 5.5\,rad. Right panel: model fits of
   the trajectories for knot-c, -h, -i, -j and -k in the innermost jet regions. 
     Within $X_n$$\sim$0.3--0.5\,mas  all the trajectories can be well
     fitted in terms of the precessing common trajectory pattern.}
     \end{figure*}
     \begin{figure*}
     \centering
    \includegraphics[width=5.6cm,angle=-90]{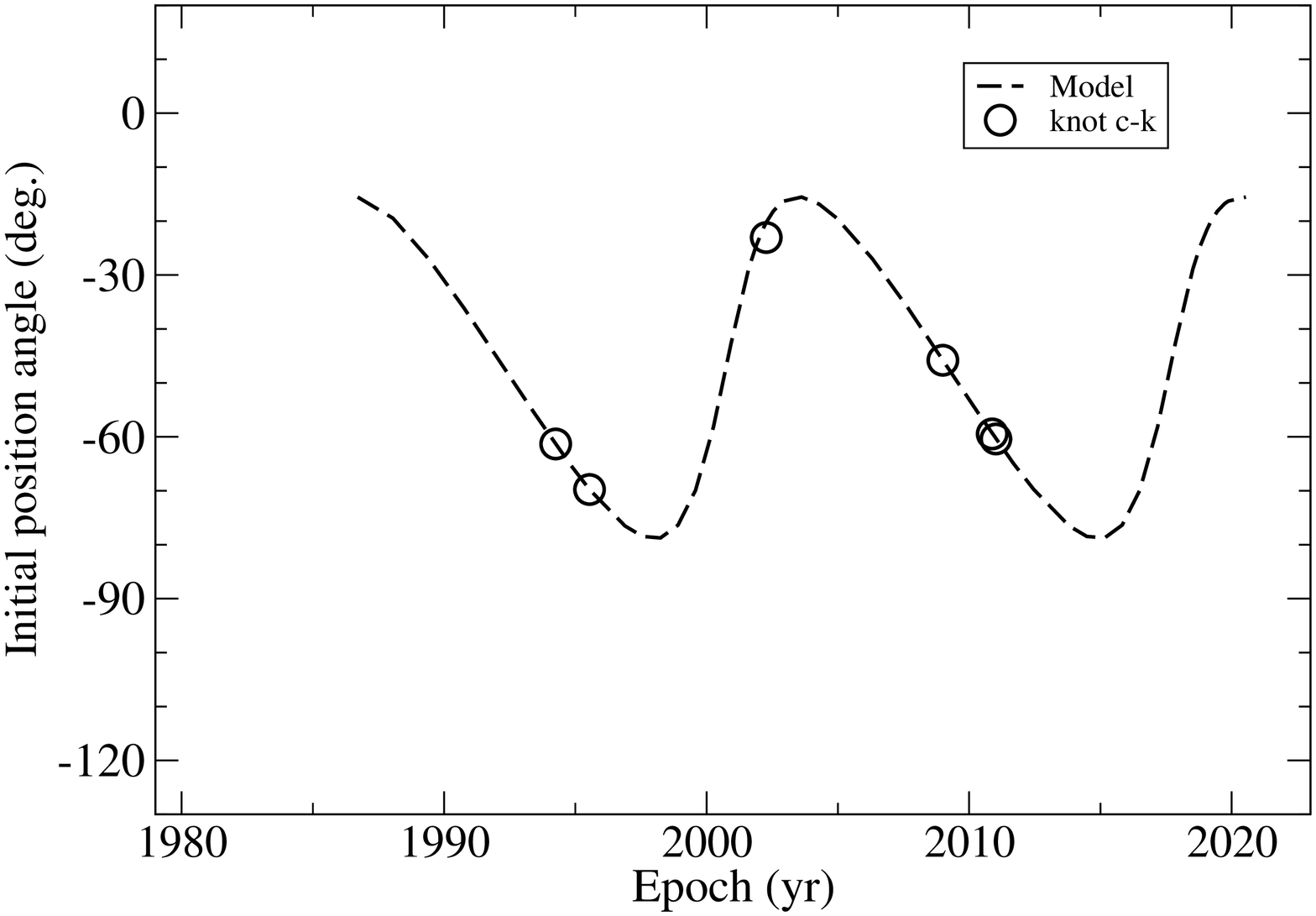}
    \includegraphics[width=5.6cm,angle=-90]{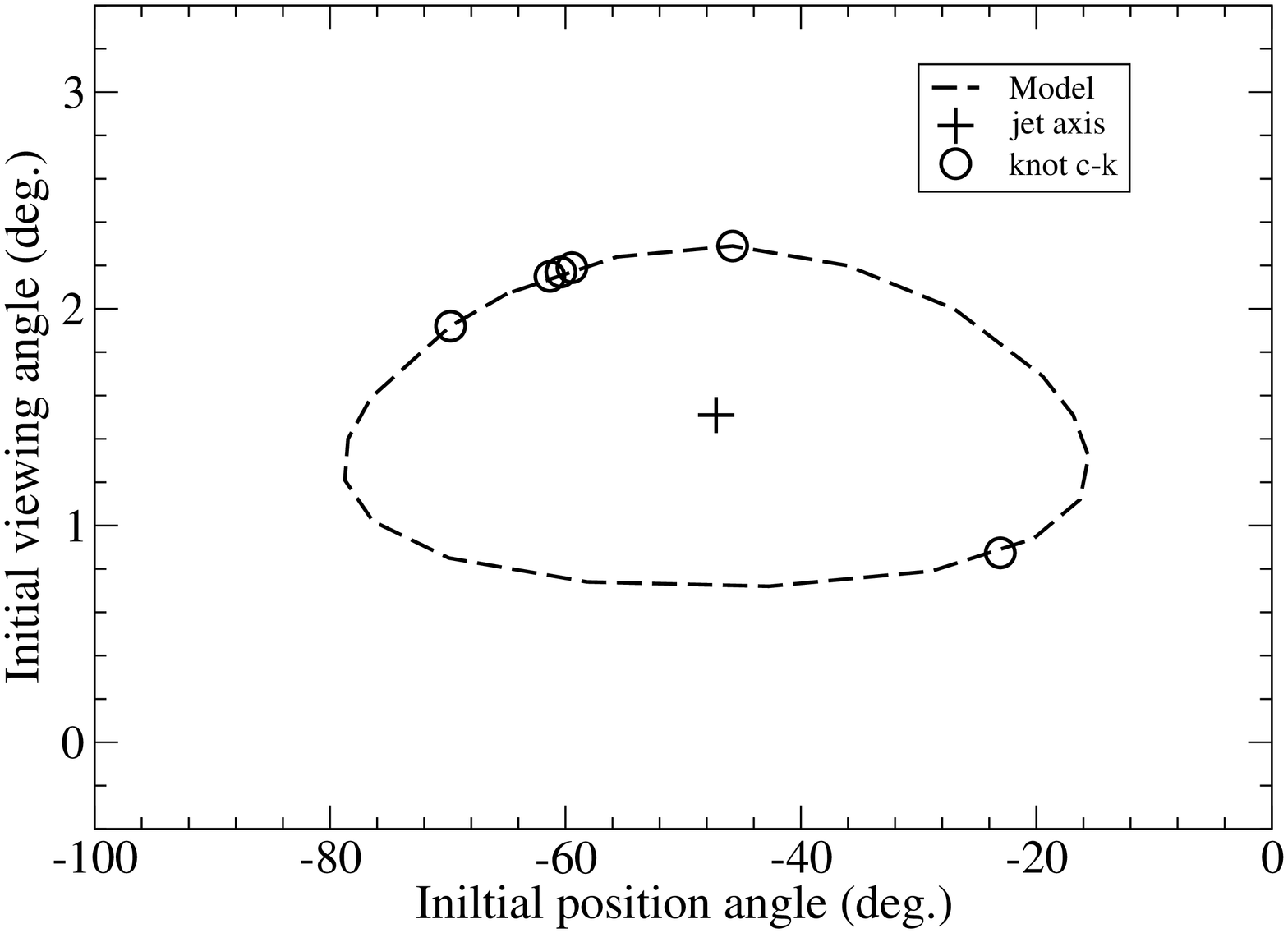}
    \caption{Left panel: the periodic swing of ejection position angle PA(t) 
     produced by the precession of the jet-nozzle for knot-c, -d, -h, -i,
     -j and -k. Right  panel: relation $\theta(PA)$
    between the ejection position angle and the viewing angle
     for these components.}
    \end{figure*}
    \section{Recapitulation of the precessing nozzle scenario}
   \begin{figure*}
   \centering
   \includegraphics[width=8.5cm,angle=0]{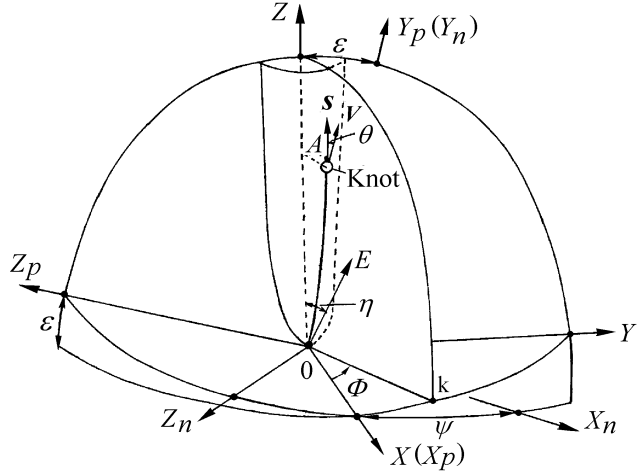}
   \caption{Geometry of the precession nozzle  model (not to scale), adopted 
   from Qian (\cite{Qi17}). Parameters $\epsilon$ and $\psi$ define 
    the (X,Z)-plane relative to the  coordinate system ($X_n, Y_n, Z_n$)
    with Z-axis taken as the axis of precession. The $Y_p$ axis indicates
    the direction toward the observer; the position of a superluminal component 
   is defined  by parameters (A, $\Phi$): $A(Z)$ is the amplitude function
  and $\Phi$ is the azimuthal angle (or phase); the plane  ($X_p$,$Z_p$)
   represents the plane of the sky; $\theta$ denotes the
   angle between the knot's velocity vector (${\bf{S}}$) 
   and the direction toward the 
   observer (${\bf{V}}$). $\eta$ is the initial half opening angle of the 
    precessing jet cone.}
   \end{figure*} 
  In order to perform  the model-fitting of the kinematics of
   B1308+326 we will use the precessing nozzle scenario previously proposed 
   in Qian et al. (\cite{Qi17}). Firstly, we recapitulate the formalism of 
   the model. The geometry of this model is shown in Figure 3, where
    three coordinate systems are used.\\
    The coordinate system 
   (${X_p},{Y_p},{Z_p}$) has the $Y_p$-axis directed toward the observer,
   i.e. the plane ($X_p$, $Z_p$) is defined as the sky plane. In this plane
   the $Z_n$-axis is defined as the  direction toward the north pole and
   the $X_n$-axis as opposite to the direction of right ascension. 
   The observed position angle of VLBI knots is measured in clockwise from the 
   $Z_n$-axis. We define a third coordinate system $(X,Y,Z)$: the $X$-axis
    coincides with the axis $X_p$ and the $Z$-axis is situated in the
   $Y_p$-$Z_p$ plane forming an angle $\epsilon$  with the $Y_p$-axis. 
   The $Z$-axis is defined as the jet-axis around  which the  precessing 
   nozzle rotates. $\psi$ denotes the 
    angle between the $X(X_p)$-axis and the $X_n$-axis.
    The precession cone has an initial half opening angle of $\eta$.\\
    We assume that the superluminal knots move along curved 
    trajectories (as shown in Figure 3), defined by  the amplitude function
    A(Z) and phase $\Phi$, which changes successively due to the nozzle 
    precession. In Figure 3, ${\bf{S}}$ denotes the direction of the spatial
    velocity and ${\bf{V}}$ denotes the 
   direction toward the observer (parallel to the direction $Y_p$).
    $\theta$ denotes the viewing angle of the knot's motion.\\
     Thus, the trajectory of a knot can be described  in the $(X,Y,Z)$
    system as follows.
    \begin{equation}
    X({Z},\Phi)=A({Z}){\cos{\Phi}},
    \end{equation}
    \begin{equation}
    Y({Z},\Phi)=A({Z}){\sin\Phi}.
    \end{equation}
   The projection of the spatial trajectory on the sky plane is defined by
   \begin{equation}
  {X_n({Z},{\Phi})}={X_p({Z,\Phi})}{\cos{\psi}}-{Z_p({Z,\Phi})}{\sin{\psi}},
    \end{equation}
    \begin{equation}
    {Z_n({Z},{\Phi})}={X_p({Z,\Phi})}{\sin{\psi}}+{Z_p({Z,\Phi})}{\cos{\psi}},
    \end{equation}
    where $\psi$ is the angle between the $X(X_p)$-axis and 
    the $X_n$-axis,
    \begin{equation}
       {X_p(Z,\Phi)}=X(Z,\Phi),
    \end{equation}
    \begin{equation}
     {Z_p(Z,\Phi)}={Z}{\sin{\epsilon}}-{Y({Z},\Phi)}{\cos{\epsilon}}.
    \end{equation}
   We give the formulas for calculating the viewing angle $\theta$, Doppler
   factor $\delta$, apparent transverse velocity $v_{app}$ and elapsed time $T$ 
   after ejection as follows.
   \begin{itemize}
     \item Viewing angle $\theta$
      \begin{equation}
       {\theta}=\arccos[\cos{\Delta}(\cos{\epsilon}+
                      \sin{\epsilon}\tan{{\Delta}_p})].
      \end{equation}
     Where
     \begin{equation}
     {\Delta}=\arctan\left[\left(\frac{dX}{dZ}\right)^2+
                \left(\frac{dY}{dZ}\right)^2\right]^{1/2},
     \end{equation}
     $\Delta$ is the angle between the spatial velocity vector and 
    the $Z$-axis,
     and 
     \begin{equation}
      {{\Delta}_p}=\arctan\left(\frac{dY}{dZ}\right)
     \end{equation}
     is the projection of $\Delta$ on the $(Y,Z)$-plane.
    \item Apparent transverse velocity $v_{app}$ and Doppler factor $\delta$
     \begin{equation}
      {v_{app}}=c{{\beta}_{app}}={\frac{c{\beta}{\sin{\theta}}}
                                {1-{\beta}{\cos{\theta}}}},
     \end{equation}
     and
     \begin{equation}
      {\delta}={\frac{1}{{\Gamma}(1-{\beta}{\cos{\theta}})}},
     \end{equation}
     where $\beta$=$\frac{v}{c}$, $v$ is the spatial velocity of the knot and
     $\Gamma$=$(1-{\beta}^2)^{-1/2}$ is  the bulk Lorentz factor.
  \item Elapsed time $T$, at which the knot reaches {\bf{the}} axial 
   distance $Z$:
     \begin{equation}
     {T}={\int_{0}^{Z}}{\frac{1+z}
            {{\Gamma}{\delta}{v}{\cos{{\Delta}_s}}}}{d{Z}},
     \end{equation}
     where $z$ is the redshift of B1308+326,
     \begin{equation}
    {\Delta}_s=\arccos\left[\left(\frac{dX}{dZ}\right)^2+
                  \left(\frac{dY}{dZ}\right)^2+1\right]^{-1/2},
     \end{equation}
    where ${\Delta}_s$ is the instantaneous angle between the velocity
     vector and the $Z$-axis.
     \end{itemize}
    All coordinates and the amplitude $A(Z)$ are measured in units 
   of milliarcsecond (mas).
   $\theta$ and {\it{v}} are instantaneous quantities at an elapsed time $T$.
  \subsection{Precessing common trajectory pattern}
    As defined above, each of the superluminal components moves along a curved
   (collimated) track described by the amplitude function A(Z) and a constant
   phase $\Phi$, while for the successive knots the phase changes due to 
   precession. That is, the superluminal components move along the precessing
   common trajectory. We choose the following pattern for describing the 
   precessing common trajectory.\\
   Its amplitude $A(Z)$ is taken  as a function of $Z$:\\ 
   when {\bf{$Z$}}${\leq}b$,
   \begin{equation}
      {A(Z)}={A_0}{\frac{2b}{\pi}}
                         {\sin\left(\frac{{\pi}{Z}}{2b}\right)},
   \end{equation}
   and when $Z>b$,
   \begin{equation}
     {A(Z)}={A_0}{\frac{2b}{\pi}}.
    \end{equation}
   Parameter $b$ may be regarded as a 'collimation parameter' to describe
  the shape of the jet collimation. The phase $\Phi$ is defined  
   by parameter $\phi$ for a specific  trajectory:\\
   \begin{equation}
     {\Phi}={\phi} + {{\Phi}_0},
   \end{equation}
   ${\Phi}_0$ is an arbitrary constant  and $\phi$
    is defined as  the precessing phase.
   \begin{figure*}
   \centering
   \includegraphics[width=5.6cm,angle=-90]{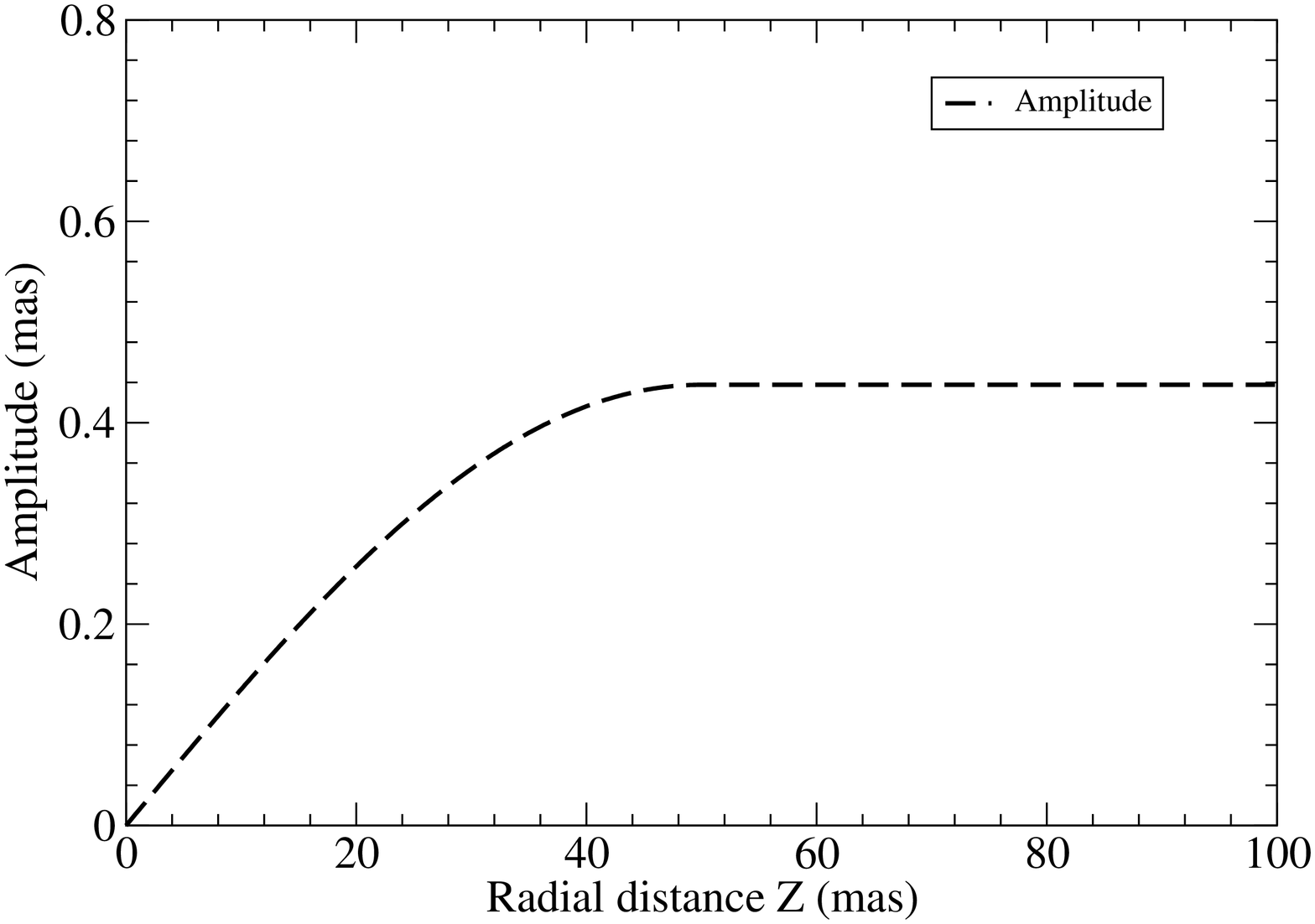}
  \includegraphics[width=5.6cm,angle=-90]{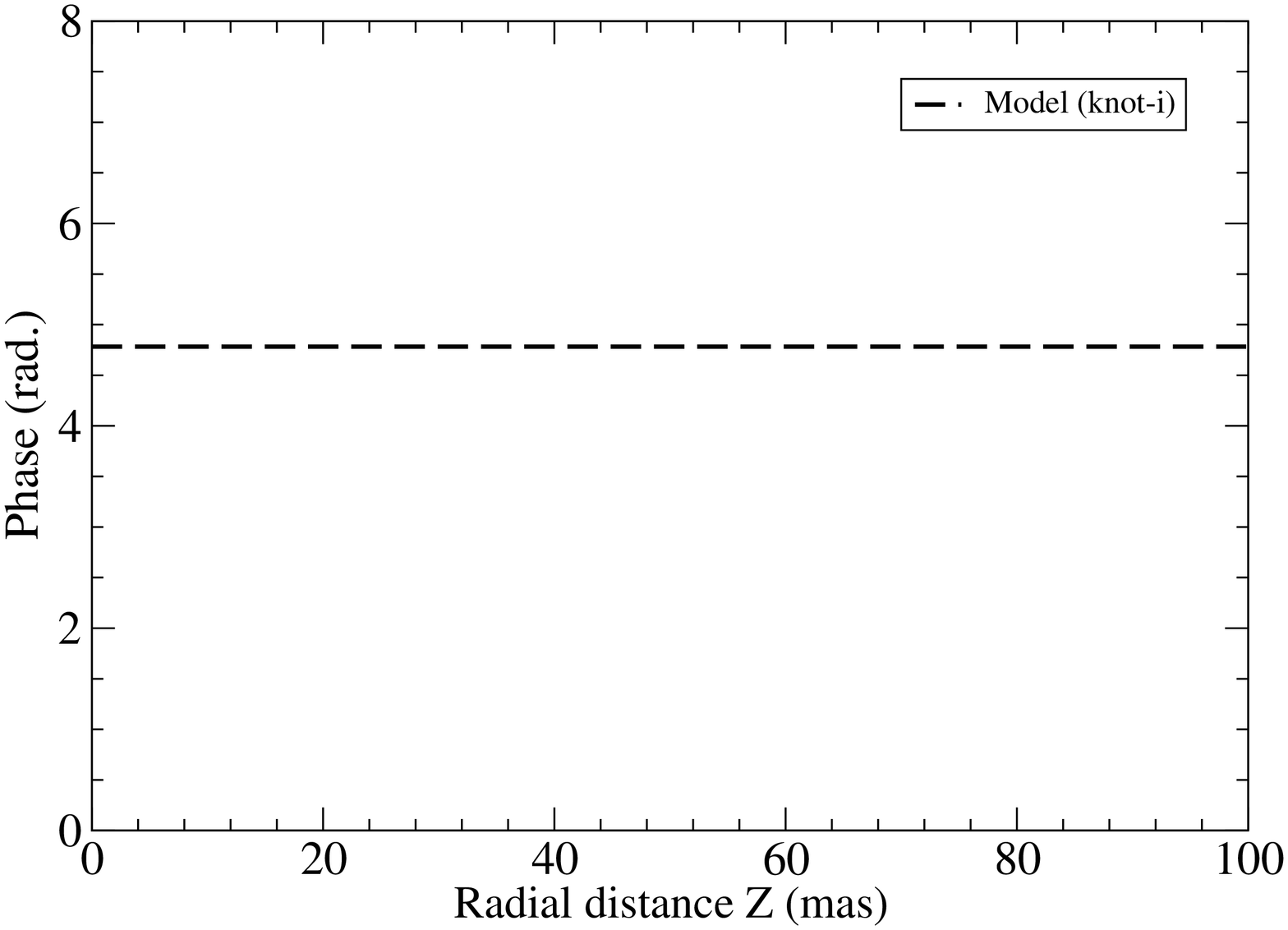}
    \caption{The assumed amplitude function $A(Z)$ and phase  $\Phi$
    for describing 
   the trajectory of a knot. Here $\Phi$=$\Phi_0$+$\phi$=4.783\,rad,
    corresponding to the precession phase $\phi$=1.0\,rad and 
    ${\Phi}_0$=3.783\,rad. See text.}
   \end{figure*}
  Since ${\frac{d\Phi}{dZ}}$=0, we have
   \begin{equation}
   {\frac{dX}{dZ}}={\frac{dA}{dZ}}{\cos{\Phi}},
   \end{equation}
   \begin{equation}
   {\frac{dY}{dZ}}={\frac{dA}{dZ}}{\sin{\Phi}}.
   \end{equation}
   Thus, from Eqs. (8), (9) and (13) we have
   \begin{equation}
   {\Delta}={\arctan\left(\frac{dA}{dZ}\right)},
   \end{equation}
   \begin{equation}
   {{\Delta}_p}={\arctan\left(\frac{dA}{dZ}\sin{\Phi}\right)},
   \end{equation}
   \begin{equation}
   {{{\Delta}_s}}=\arccos\left[\left(1+
               \left(\frac{dA}{dZ}\right)^2\right)^{-1/2}\right].
   \end{equation}
   Substituting $\Delta$, ${\Delta}_p$ and ${\Delta}_s$ into
    Eqs. (7) and (10)--(12), we can calculate
 the viewing angle $\theta$, apparent velocity ${\beta}_{app}$,
   Doppler factor $\delta$ and elapsed time $T$.\\
  We should point out that the assumed pattern of the precessing 
   common trajectory (see Figure 4 below) closely
    represents the field structure configurations observed in the jets of 
  radio galaxies and blazars.
   For example, the giant radio galaxy M87 which has a powerful optical-radio
   jet and a supermassive black hole of $\sim$6$\times$$10^9{M_{\odot}}$,
   is the best possible target for studying the initial
   jet formation/collimation process (Biretta et al. \cite{Bi02}).
    Nakamura \& Asada (\cite{Na13}) (also, Asada \& Nakamura \cite{As12};
    Doeleman et al. \cite{Do12}) have found that its 
   innermost jet emission components follow an extrapolated parabolic 
   streamline, so that  the jet has a single power-law structure with a nearly
   five orders of magnitude in the distance starting 
    from the vicinity of the supermassive black hole, less 
   than ten Schwarzschild radii.
   They have also proposed a magnetohydrodynamic nozzle model to interpret the
    property of the bulk jet acceleration and assumed that the MHD nozzle
   consists of a hollow parabolic tube. Most recently, Lu et al. (\cite{Lu23})
   (also cf. Kim et al. \cite{Kim23}) present the parabolic jet profile 
   extending to the jet-axis distance of $\sim$70$\mu$as 
   (or $\sim$10$R_s$, $R_s$ -- Schwarzschild radius). \\
   Moreover, general relativistic MHD simulations (e.g., McKinney et al.
   \cite{Mc12}) reveal that the magnetic field structures (or configurations)
   near the horizon of a rotating black hole closely correspond to
    a parabolic configuration which is consistent with the analytic results
    given by Beskin \& Zheltoukhov (\cite{Bes13}) for a field geometry:
   a radial field near the horizon and a vertical field far from the black hole.
    In these configurations, the distribution of the magnetic field and the
   field angular  velocity profile near the horizon can be described
    in detail (Punsly \cite{Pu01}; McKinney et al. \cite{Mc12};
    Beskin \& Zheltoukhov \cite{Bes13}). 
   In addition, the assumed pattern is also quite similar to the 
   fork-structure observed in the prominent blazar OJ287 by Tateyama 
   (\cite{Ta13}). \\
     In fact, in the previous works we have already adopted such a kind 
    of common precessing trajectory  pattern to study the kinematics of the
    superluminal components in a few blazars, e.g., 3C279 (Qian et al.
    \cite{Qi19}; Qian \cite{Qi12}, 
   \cite{Qi13},), 3C454.3 (Qian et al. \cite{Qi14}, \cite{Qi21}), OJ287 (Qian
   \cite{Qi18b}), 3C345 (Qian et al. \cite{Qi91}, Qian \cite{Qi22a}, 
    \cite{Qi22b}), and also in the QSOs PG1302-132 (Qian et al. \cite{Qi18a})
    and NRAO 150 (Qian \cite{Qi16}).\\
   In this paper we will adopt the concordant cosmology model with
   ${\Omega}_{\Lambda}$=0.73, ${\Omega}_m$=0.27 and  Hubble 
   constant $H_0$=71 km\,${{\rm{s}}^{-1}}$\,${{\rm{Mpc}}^{-1}}$ 
   (Spergel et al. \cite{Sp03}). For the redshift z=0.997 of B1308+326, 
    we have its luminosity distance $D_L$=6.61\,Gpc and 
    the angular diameter
   distance $D_A$=1.66\,Gpc (Hogg \cite{Ho99}; Pen \cite{Pe99}). 
    The angular scale is
    1\,mas=8.04\,pc and the proper motion of 1\,mas/yr is
     equivalent to an apparent velocity of 52.34c.
   \section{Selection of model parameters}
    In our precessing jet
    nozzle model, the jet nozzle  precesses around a fixed jet axis and the
     knots  are ejected from the nozzle, moving along their  individual 
    trajectories (of a common pattern, ballistic or helical, Qian \cite{Qi16})
    with different bulk 
    Lorentz factors. The precession of the nozzle leads to the rotation of 
    the ejection direction of the knots or the periodic position angle swing.
    The combination of a sequence of isolated knots (and associated magnetized
    plasma flows) ejected from this nozzle will construct the structure of
    the whole jet and its evolution  seen on VLBI-maps (e.g., Tateyama 
    \& Kingham \cite{Ta04}; Qian et al. \cite{Qi17}; Tateyama \cite{Ta09}, 
    \cite{Ta13}; Qian \cite{Qi14}). \\
     In order to model-fit the kinematics of the superluminal
      knots in terms of our precessing jet-nozzle model, the model 
     parameters, defining the jet-axis direction ($\epsilon$, $\psi$),
     the pattern of the precessing common track ($A_0$, b), precession period 
     $T_p$ and phase ($\Phi_0$, $\phi$) should be  approriately 
     selected. We shall adopt the same values as used in the previous work
     (Qian et al. \cite{Qi17}) as follows:
   \footnote{Here for the superluminal knots, we shall use the changes 
   in parameters $\epsilon$ and ${\psi}$ (instead of changes in amplitude A(Z))
    to describe the transition from the common precessing tracks in the
     innermost jet regions to their own individual trajectories in the 
     outer jet regions.} \\
   \\
   $\epsilon$ = $1.5^{\circ}$ \\
   $\psi$ = --$42.8^{\circ}$ \\
   $A_0$ = 1.375$\times{10^{-2}}$\,mas \\
   $b$ = 50\,mas \\
   ${\Phi}_0$ = 3.783\,rad \\
   $T_p$ = 16.9\,yr \\
   \\
   The ejection epoch $t_0$ for the knots can be calculated  
   from their precession phase $\phi$:\\
   \begin{equation}
      {t_0}=1995.54+{\frac{T_p}{2\pi}}(6.0-{\phi}).
   \end{equation}
     The kinematic parameters including the bulk Lorentz factor, viewing angle,
    apparent velocity and Doppler factor as function of time will be derived
    through the model-fitting process.\\
     In order to model-fit the light curves of the superluminal knots
     the observed flux density ${S_{obs}}(\nu,t)$ can be calculated as:\\
   \begin{equation}
     {{S_{obs}}(\nu,t)}={{S_{int}}(\nu,t)}{[\delta(t)]^{3+\alpha(\nu,t)}}
   \end{equation}
   ${S_{int}}(\nu,t)$ -- intrinsic flux density; $\delta(t)$ -- Doppler factor;
   $\alpha(\nu,t)$ -- spectral index ($S_{\nu}$\,$\propto$\,${\nu^{-\alpha}}$).
    In most cases ${S_{int}}(\nu,t)$=${S_{int}}(\nu)$
     and $\alpha(\nu,t)$=$\alpha(\nu)$ are assumed.
   \begin{figure*}
   \centering
   \includegraphics[width=5.6cm,angle=-90]{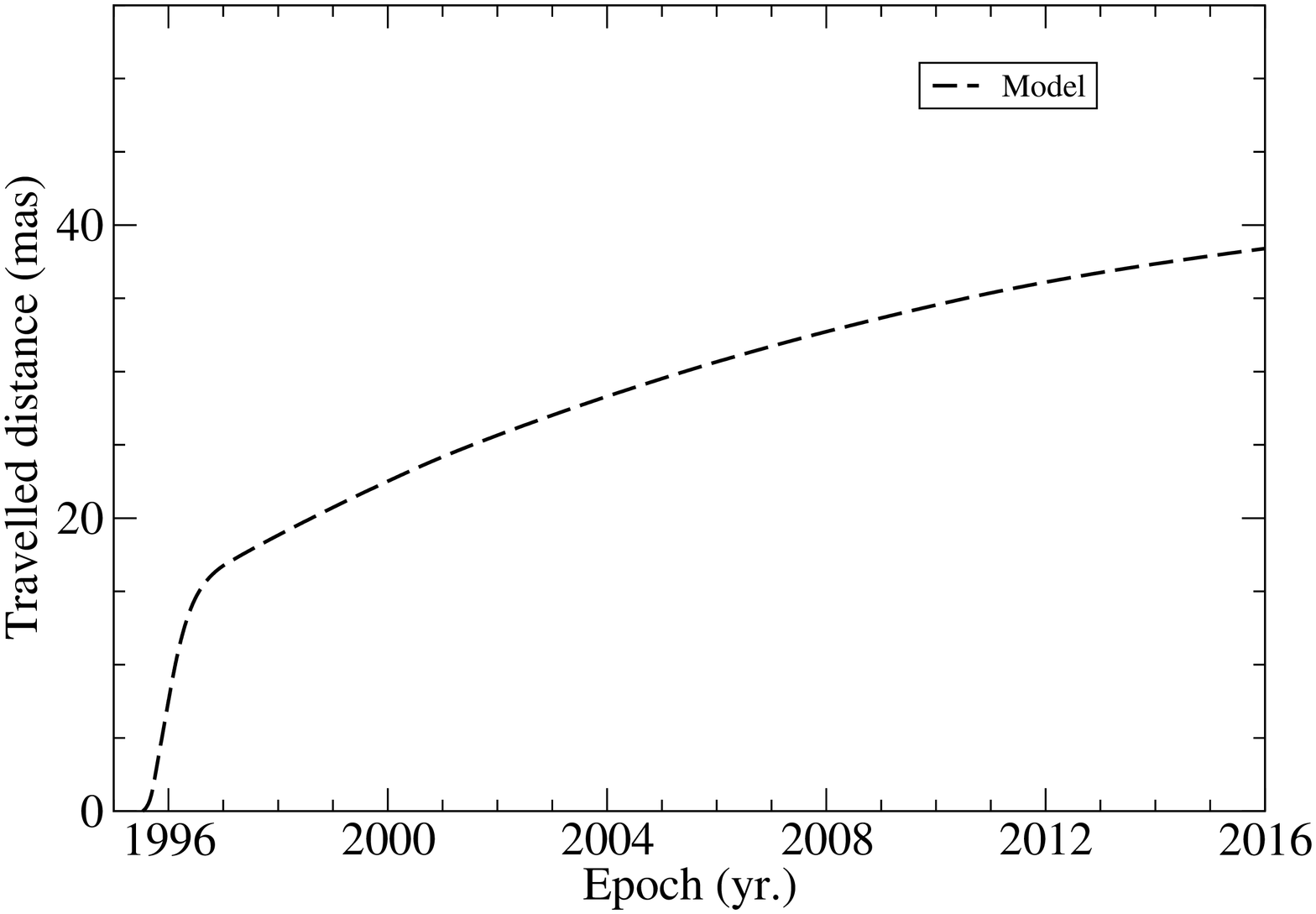}
   \includegraphics[width=5.6cm,angle=-90]{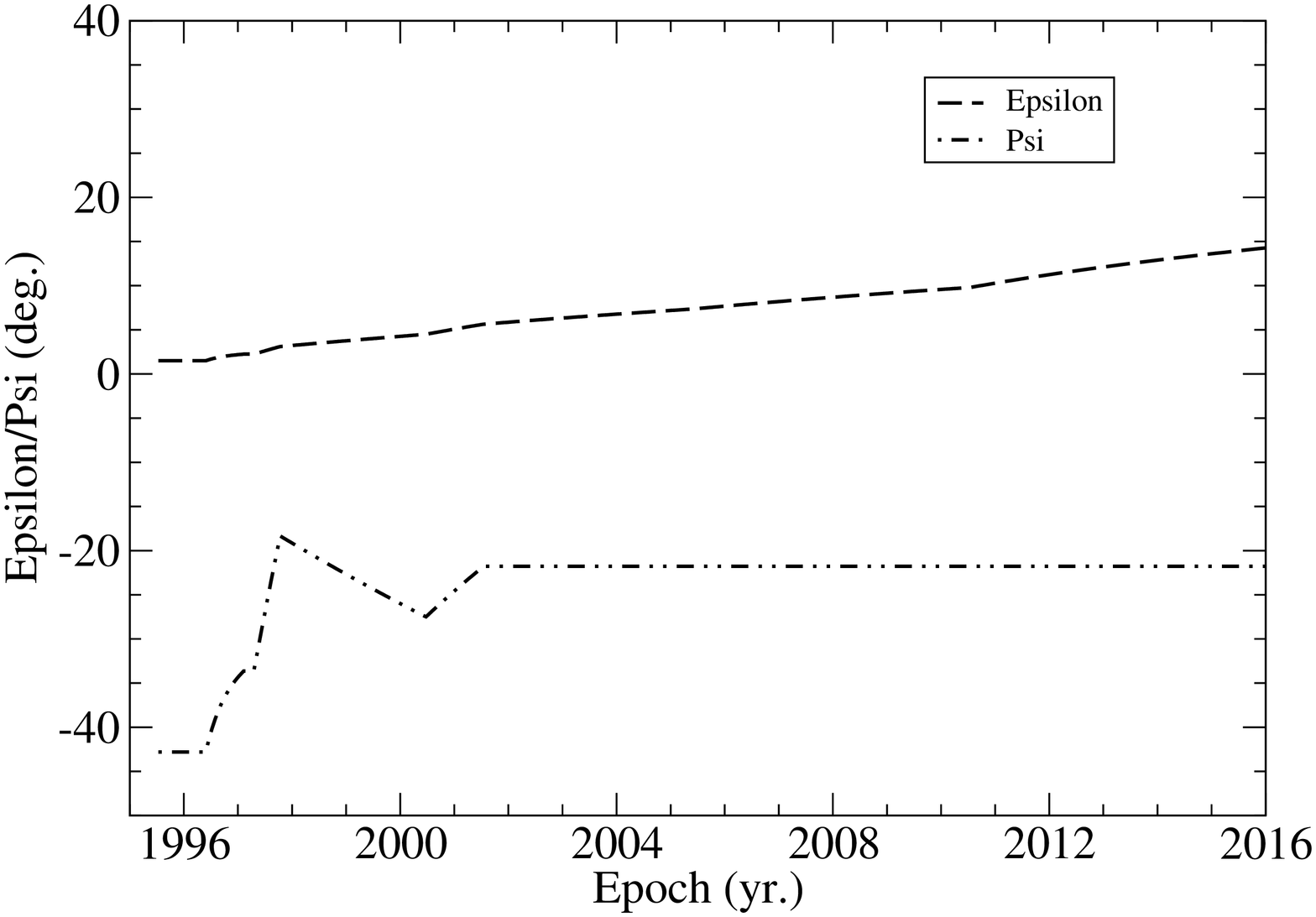}
   \caption{Knot-c: Model fitting of the entire kinematics (1995--2014). 
     Left panel: the modeled traveled
    distance Z(t) along  the Z-axis. Right panel: the modeled curves for parameters
     $\epsilon(t)$ and
   $\psi(t)$. After 1996.40 $\psi$ increased quickly, implying a rotation
    of the (X,Z)-plane relative to the coordinate system ($X_n, Y_n, Z_n$)
   and knot-c started to move along its own individual track.}
   \end{figure*}
   \begin{figure*}
   \centering
   \includegraphics[width=5.6cm,angle=-90]{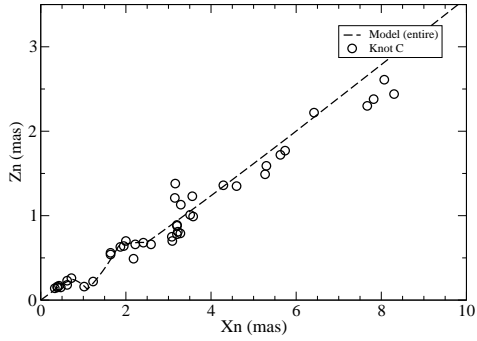}
   \caption{Knot-c: the model fit of the entire trajectory extending to core 
    distance $r_n$$\sim$8.5\,mas (during 1996--2014). The curvature within
    $X_n{\sim}$1.7\,mas is well fitted.}
   \end{figure*}
   \begin{figure*}
   \centering
   \includegraphics[width=5.6cm,angle=-90]{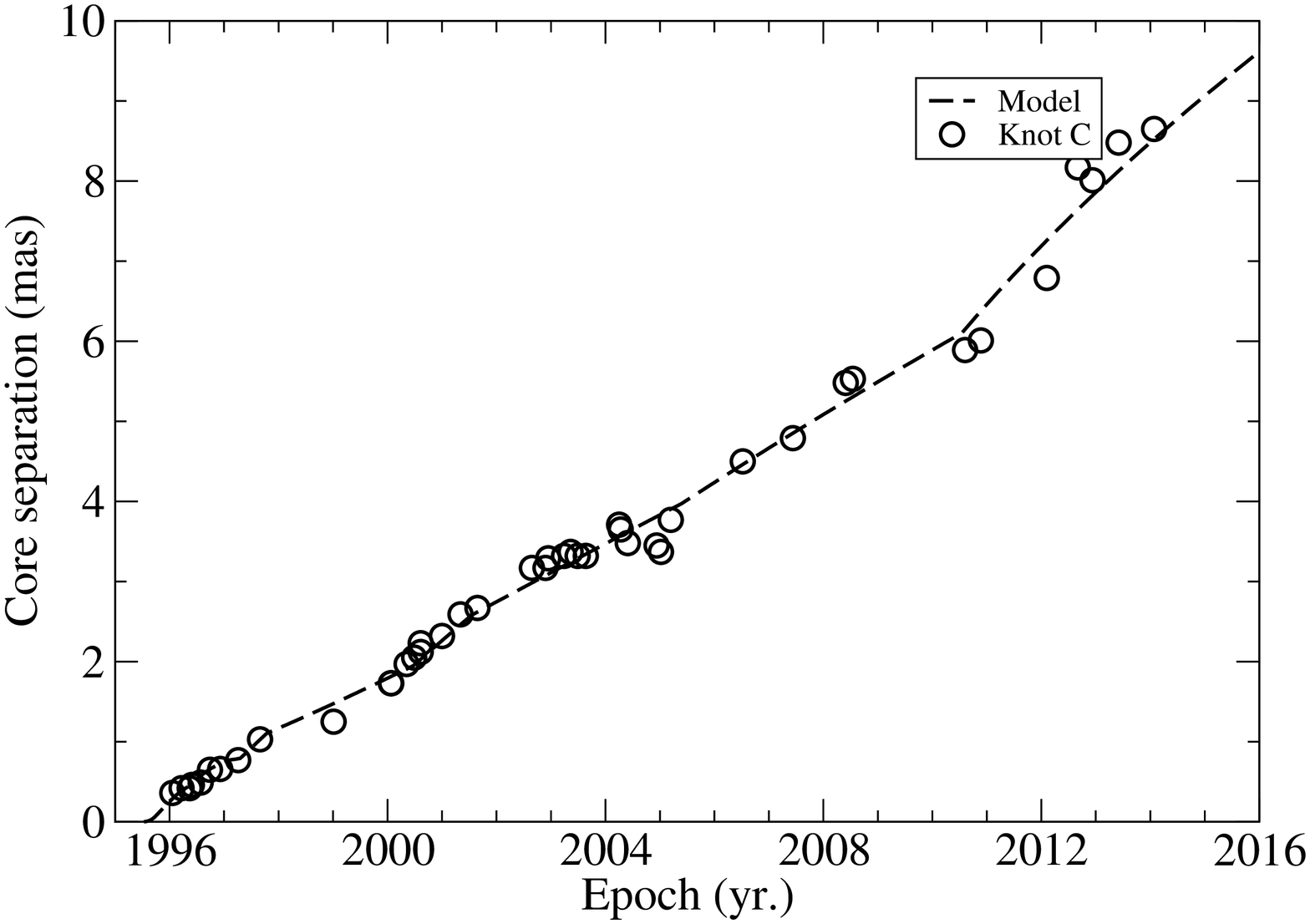}
   \includegraphics[width=5.6cm,angle=-90]{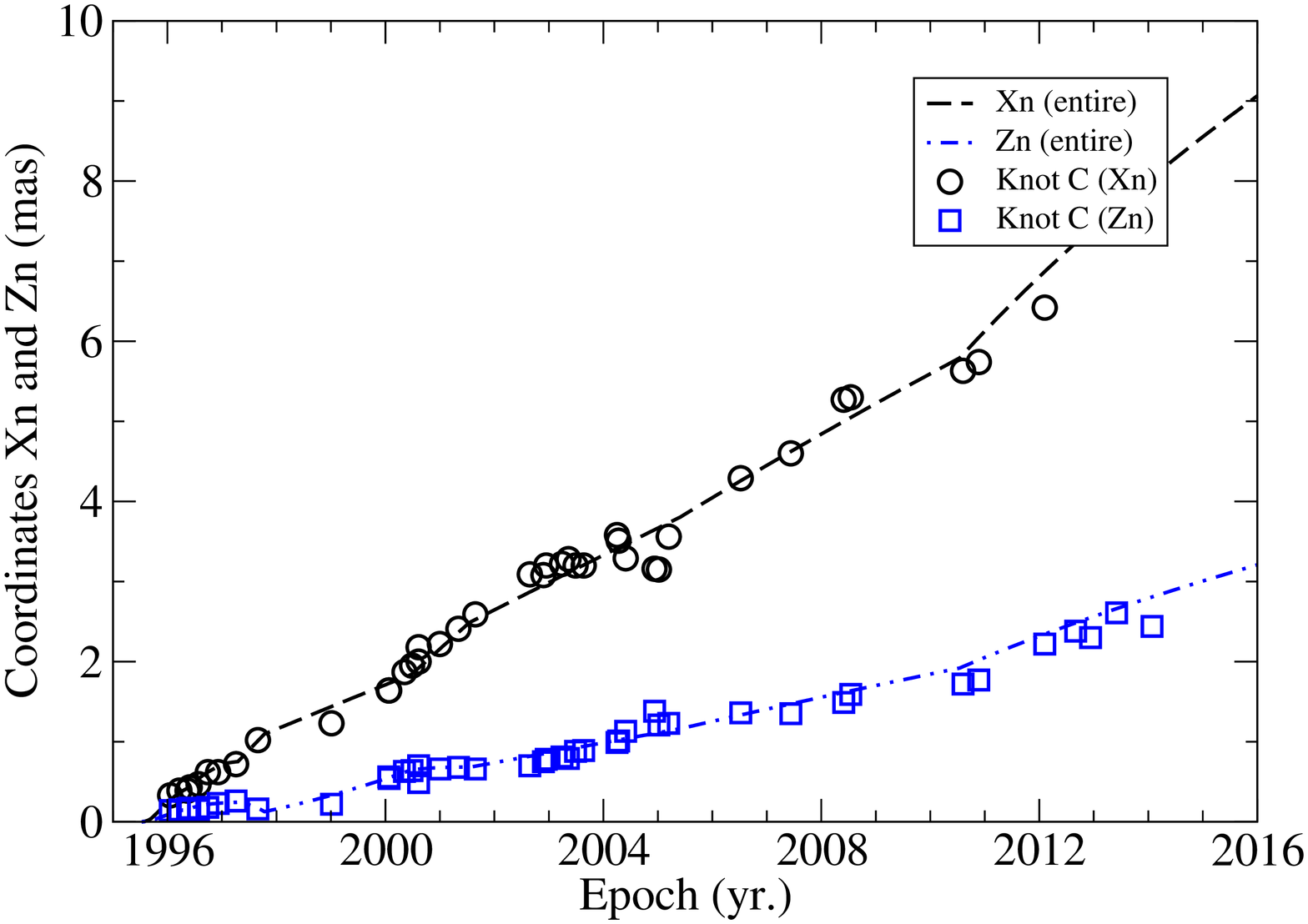}
   \caption{Knot-c: Model fitting of the entire kinematics (1995--2014).
    Left panel: the model-fit 
   of the core separation $r_n(t)$. Right panel: the model-fits of 
   coordinates $X_n(t)$ and $Z_n(t)$. All the kinematic features are well
   fitted in terms of the precessing nozzle model extending to core
   distance  $r_n{\sim}$8.5\,mas (during 1996--2014).}
   \end{figure*}
   \begin{figure*}
   \centering
   \includegraphics[width=5.6cm,angle=-90]{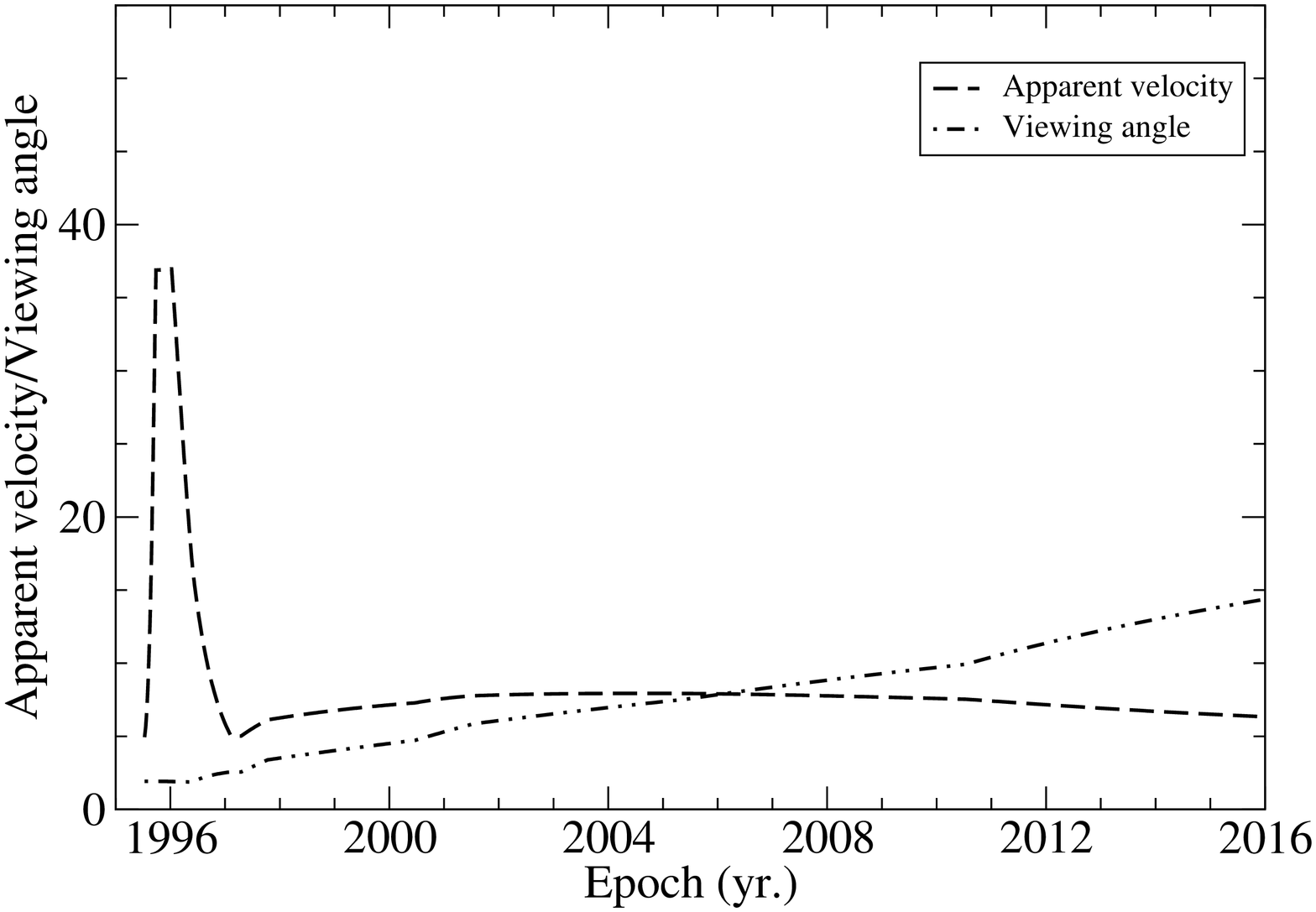}
   \includegraphics[width=5.6cm,angle=-90]{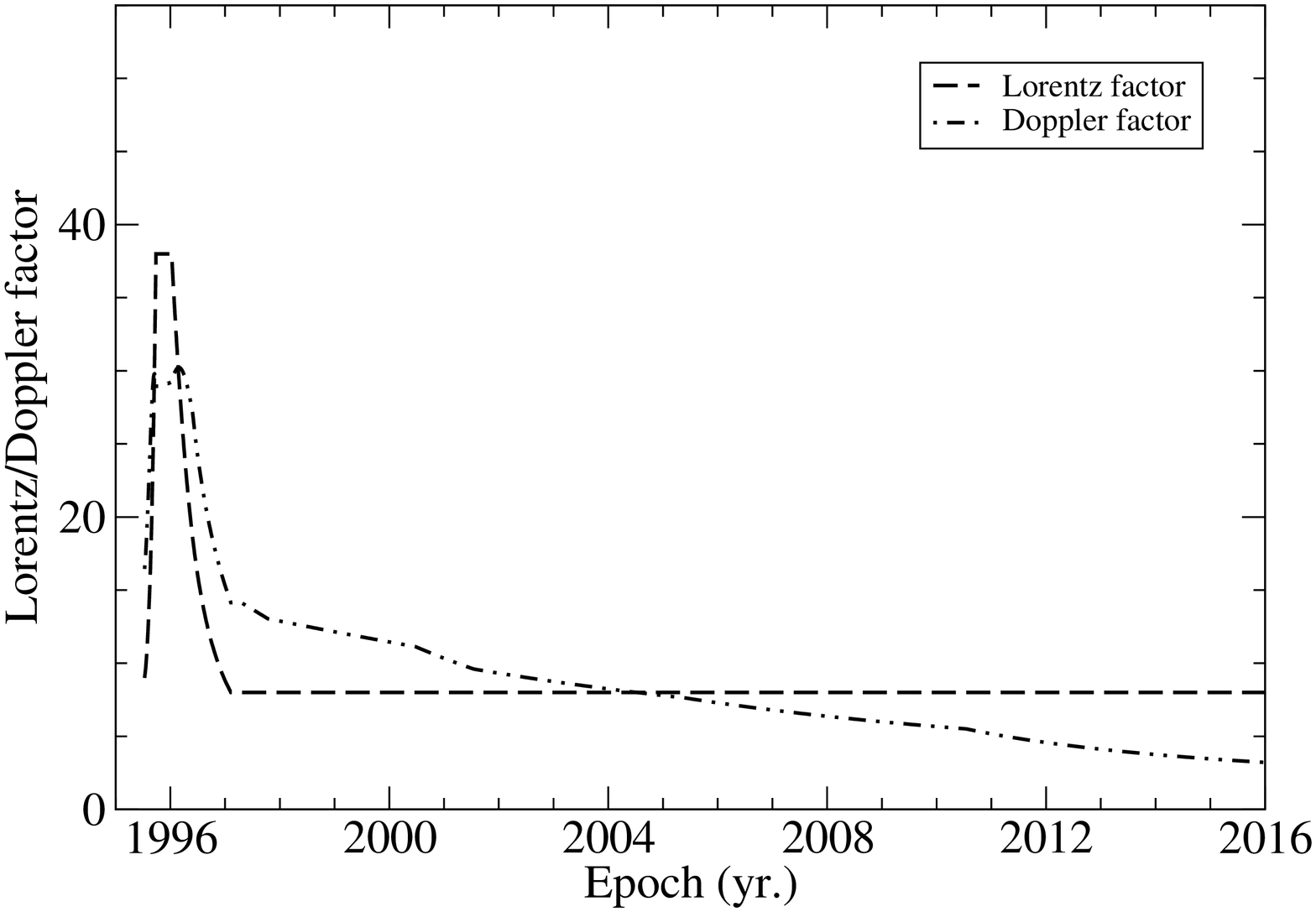}
   \caption{Knot-c: Model fitting of the entire kinematics (1995--2014).
     Left panel: the
     model-derived apparent velocity $\beta_{app}(t)$ and viewing 
   angle $\theta(t)$. Right panel: the model-derived bulk Lorentz factor 
   $\Gamma(t)$ and Doppler factor $\delta(t)$. The prominent feature is 
   $\delta(t)$$<$$\Gamma(t)$ during the peaking stage.}
   \end{figure*}     
  \section{Knot-c: Model-fitting results} 
   The model fitting results for knot-c will be presented in two parts: 
    (1) for entire kinematics (1995--2014) in Fig.5--8 and (2) for the
    inner jet region (1995--2001.5, $r_n{\leq}2.5$\,mas) in Fig.9-13. \\
  \subsection{Knot-c:  model-fitting of entire kinematics (1995-2014)}
    Its ejection epoch $t_0$=1995.54, corresponding to a precession phase 
    $\phi$=6.0\,rad.\\
    In Fig.5 the traveled distance Z(t) along Z-axis (left panel) and 
    the parameters $\epsilon(t)$ and $\psi(t)$ (right panel) are presented.
    During the time-interval $\sim$1996--2000 parameter $\psi$  changed 
    quickly, implying a rotation of the XY-plane \footnote{XY-plane is defined
    as the reference-plane for calculating the precession phase of knots.} 
    relative to the coordinate system ($X_n,Z_n$).\\
     The model-fits of the entire trajectory $Z_n(X_n)$, core separation
     $r_n(t)$, coordinates $X_n(t)$ and $Z_n(t)$ are shown in Fig.6 and Fig.7,
     respectively. Within $r_n(t)\sim$8.6\,mas (or till 2014.09) all 
     these kinematic  features were well fitted.\\
     The model-derived apparent velocity $\beta_{app}(t)$ and viewing angle 
     $\theta(t)$ (left panel) and bulk Lorentz factor $\Gamma(t)$ and Doppler
     factor $\delta(t)$ (right panel) are presented in Fig.8. Both show a peak
     structure during $\sim$1995.5--1997.0, coincident with the radio 
     burst (see Fig.13, below). 
      \begin{figure*}
     \centering
     \includegraphics[width=5.6cm,angle=-90]{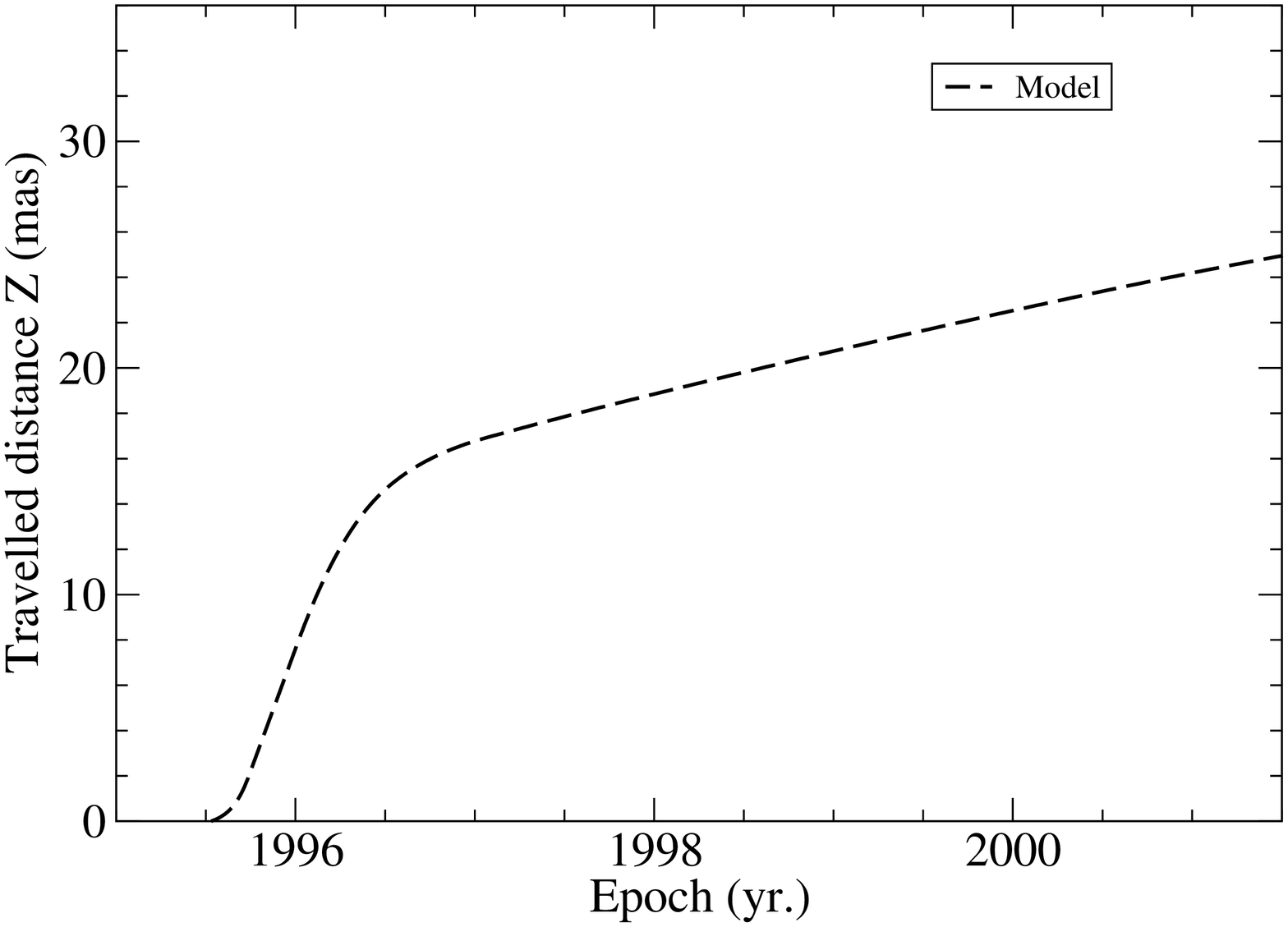}
     \includegraphics[width=5.6cm,angle=-90]{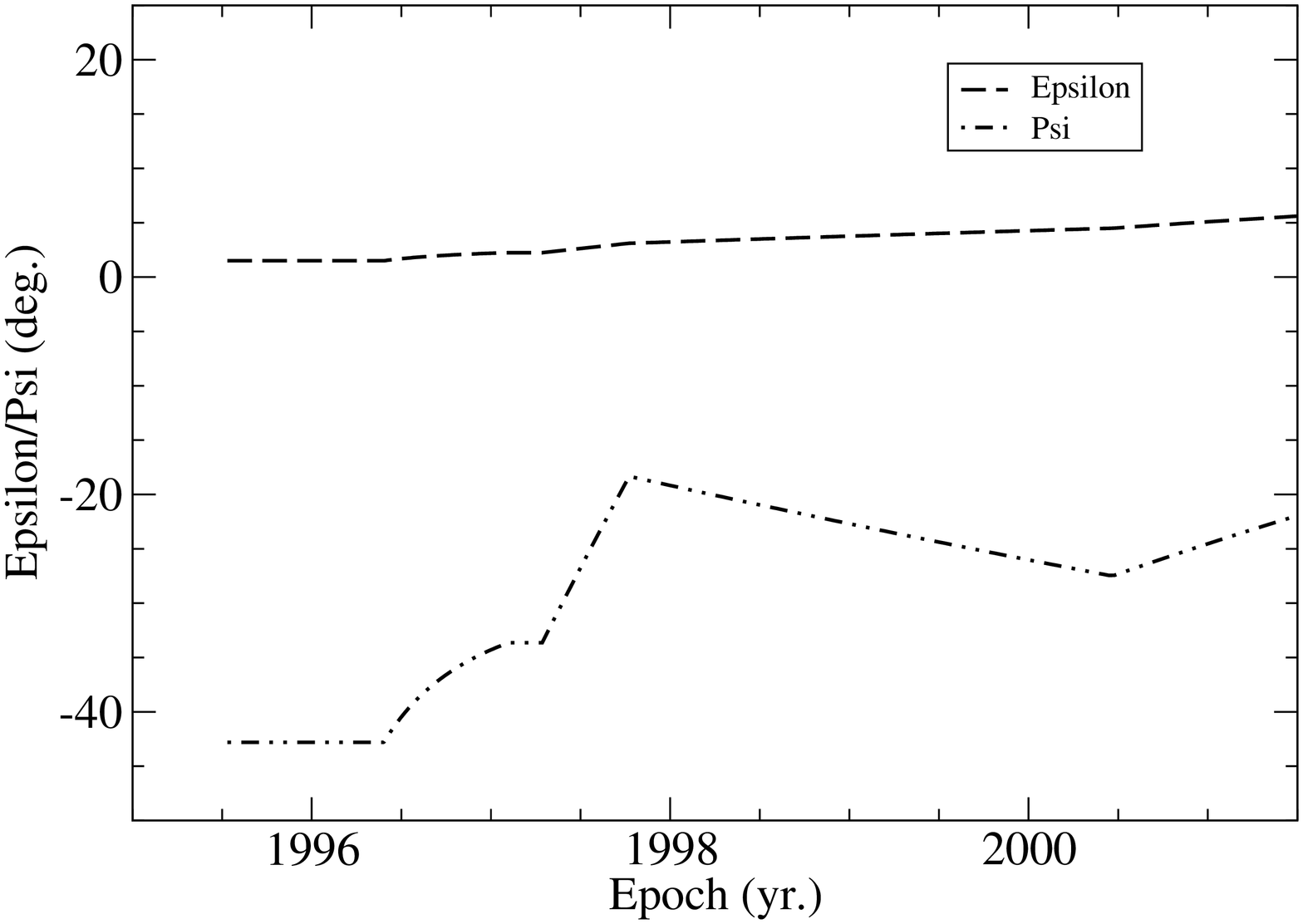}
     \caption{Knot-c: model fitting of the inner kinematics (1995.5--2001.5).
      Left panel: the modeled traveled distance Z(t) along the Z-axis. Right
      panel: the modeled curves for parameters
     $\epsilon(t)$ and $\psi(t)$. Within core separation $r_n\simeq$0.46\,mas
     (or before 1996.40, corresponding to traveled distance 
     Z$\leq$13.8\,mas=106\,pc) $\epsilon$=$1.5^{\circ}$ and 
     $\psi$=--$42.8^{\circ}$, knot-c moved along the precessing common 
    trajectory. Beyond $r_n$=0.46\,mas, 
     parameter $\psi$ increased quickly and knot-c started to move along its 
     own individual track.}
     \end{figure*}
     \begin{figure*}
     \centering
     \includegraphics[width=6.5cm,angle=-90]{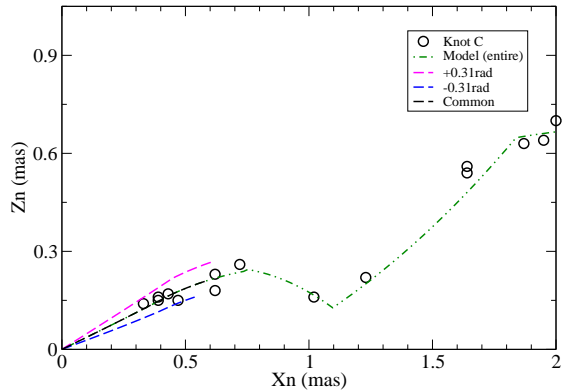}
     \caption{Knot-c: Model fitting of the inner trajectory
      ($X_n{\leq}$2\,mas, 1995.5--2001.5). Within core separation $r_n$=0.46\,mas (or coordinate
     $X_n{\leq}$0.43\,mas) knot-c moved along the precessing common trajectory.
     Beyond that knot-c started to move along its own individual trajectory. 
     The red and green curves are calculated for the precession phases 
     $\phi{\pm}$0.31\,rad ($\phi$=6.0\,rad), demonstrating that the 
     observational data-points are within the area confined by the two
     curves and showing that the precession period was derived correctly
     with an uncertainty of ${\sim}\pm$5\% of the precession period.}
     \end{figure*}
     \begin{figure*}
     \centering
    \includegraphics[width=5.6cm,angle=-90]{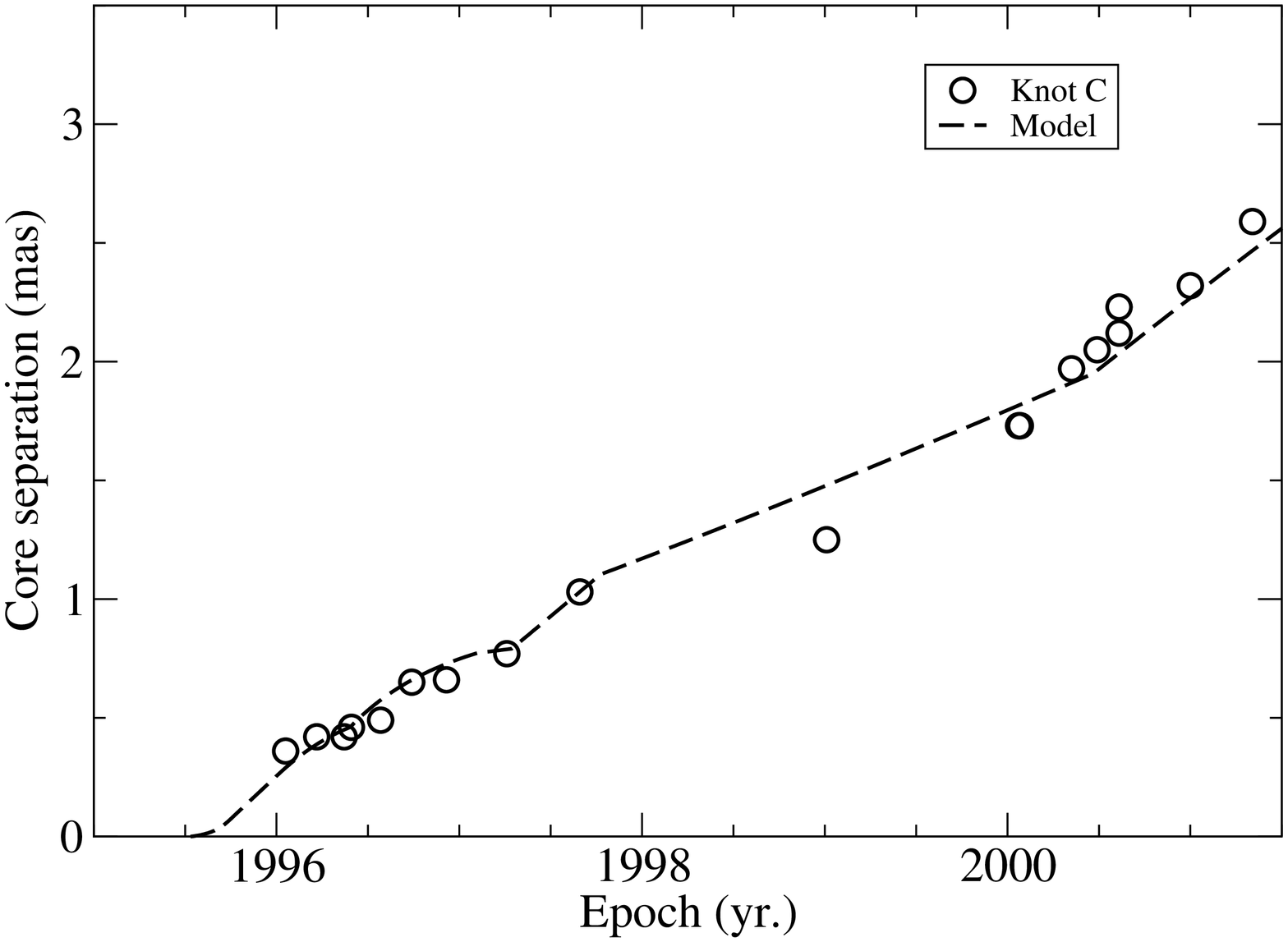}
     \includegraphics[width=5.6cm,angle=-90]{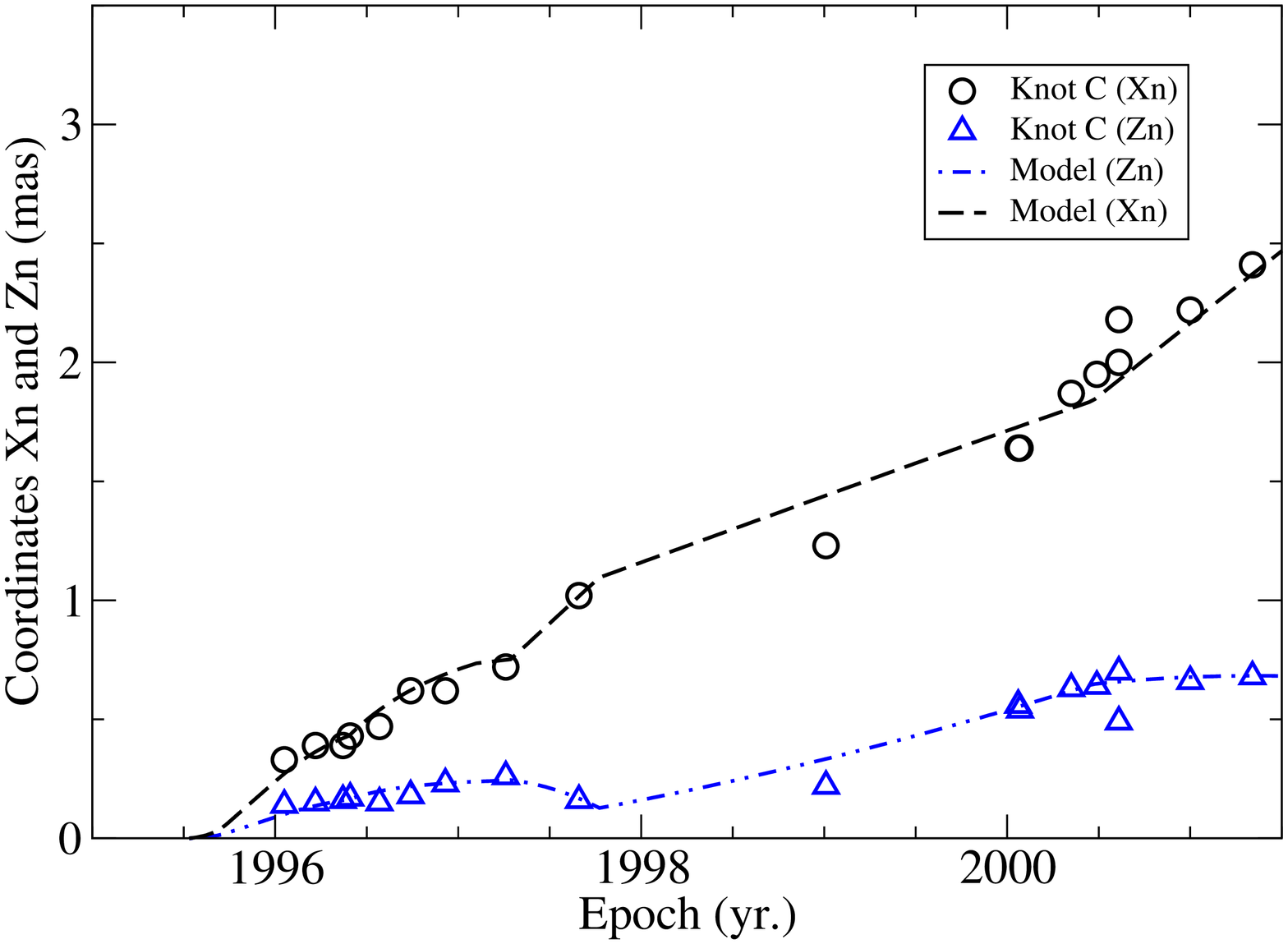}
     \caption{Knot-c: Model fitting of the  kinematics (1995.5--2001.5).
      Left Panel: the model fit of the core separation $r_n(t)$. Right panel:
      the model fits of coordinates $X_n(t)$ and $Z_n(t)$. All the kinematic 
     features are well fitted extending to $r_n{\sim}$ 2.5\,mas. Before 1996.40
     knot-c moved along the precessing common traejctory, while after that it
     started  to move along its own individual track.}
     \end{figure*}
     \begin{figure*}
     \centering
     \includegraphics[width=5.6cm,angle=-90]{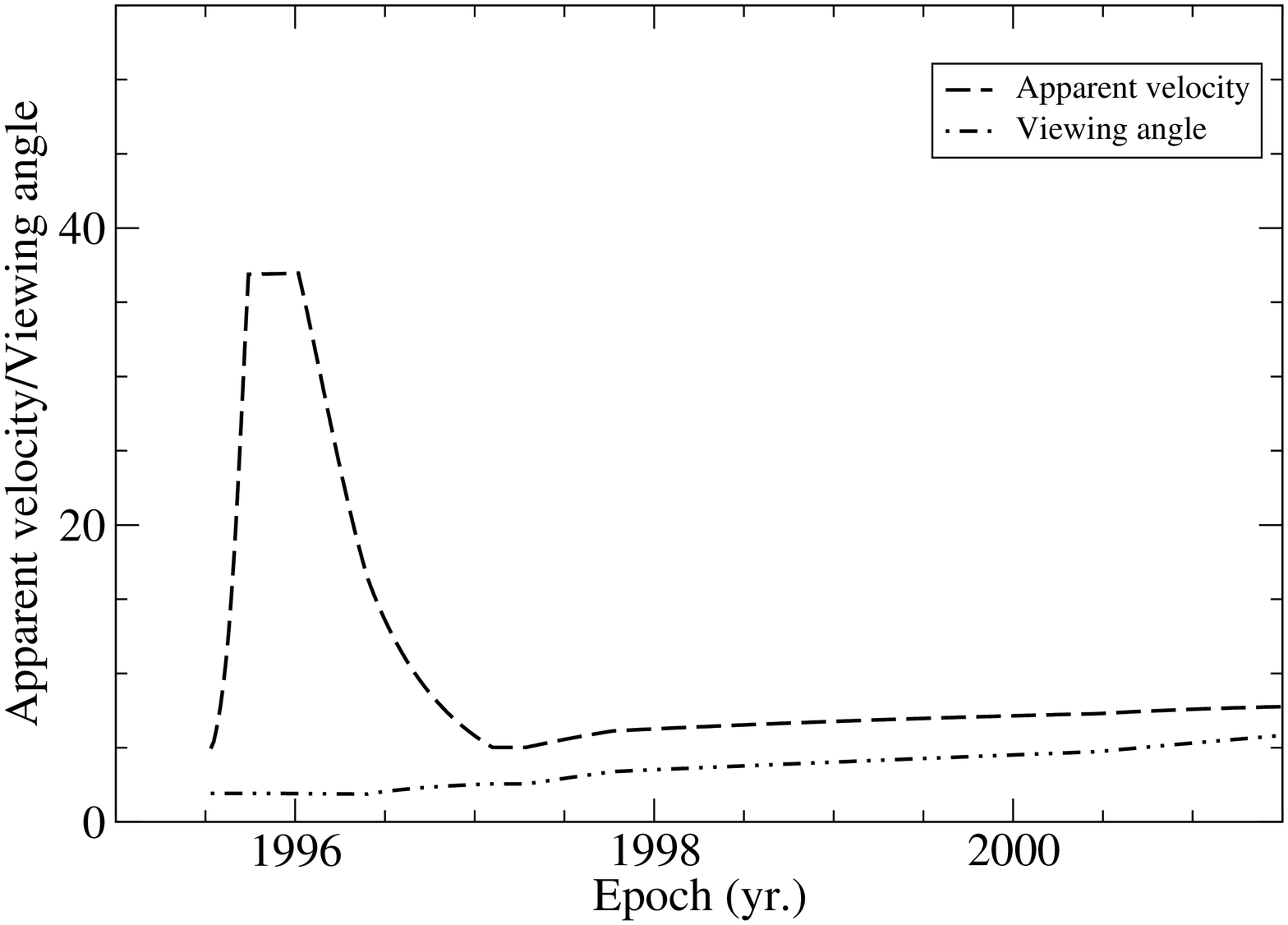}
     \includegraphics[width=5.6cm,angle=-90]{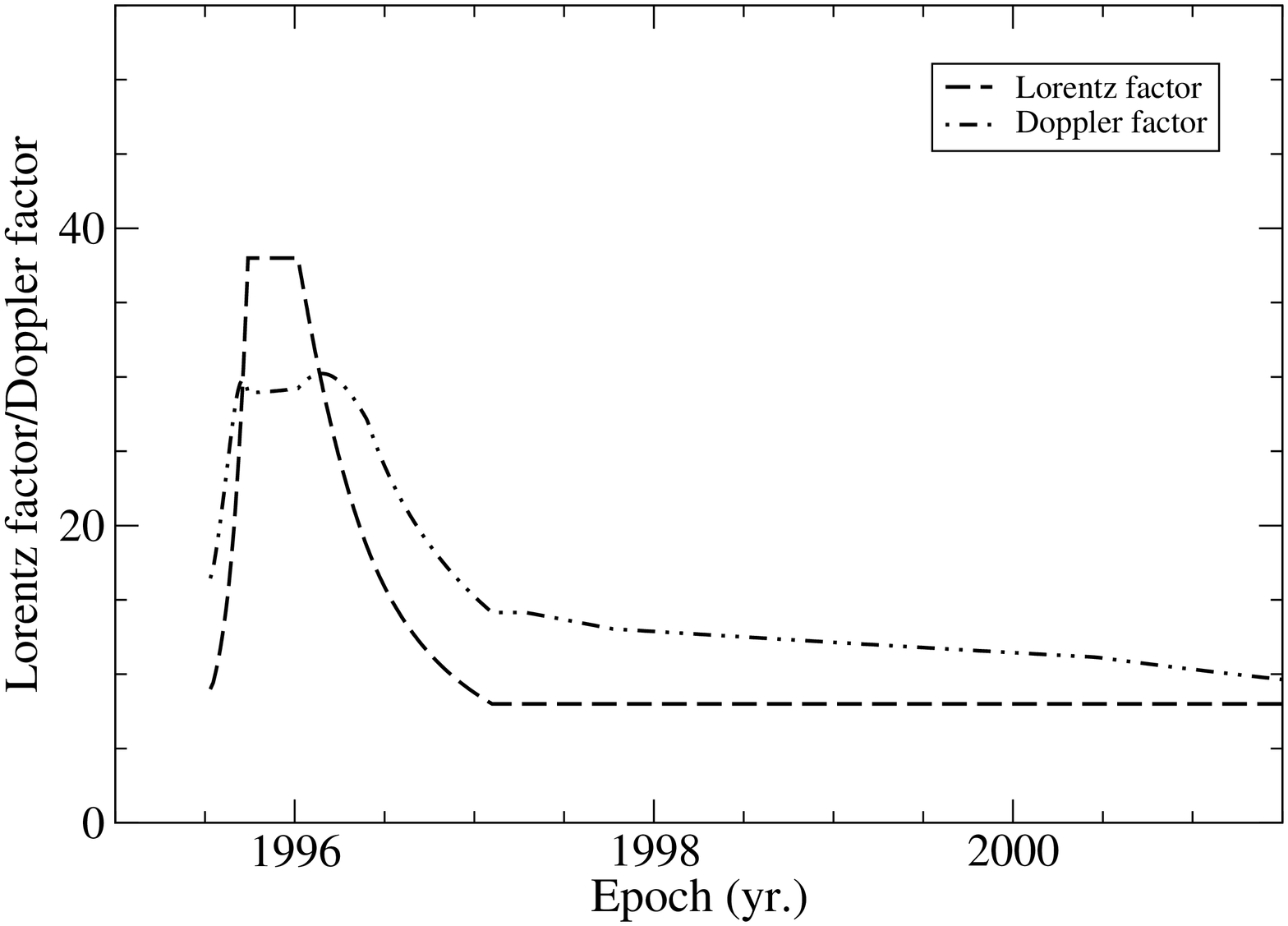}
     \caption{Knot-c: model fitting of the inner kinematics (1995.5--2001.5).
     Left panel: the
      model-derived apparent velocity $\beta_{app}(t)$ and viewing angle
      $\theta(t)$. Right panel: the model-derived bulk Lorentz factor
      $\Gamma(t)$ and Doppler factor $\delta(t)$.
       At 1996.15 $\delta$=$\delta_{max}$=30.2, $\Gamma$=29.7, $\beta$=29.6,
       while at 1996.02 $\Gamma$=$\Gamma_{max}$=38.0
       and $\beta_{app}$=$\beta_{a,max}$=37.0. $\theta(t)$ varied in the range 
     of [$1.92^{\circ}$, $5.32^{\circ}$] during 1995.5--2001.0. It is worthy of
     note that both $\Gamma(t)$ and $\delta(t)$ have a bump structure and 
     during the peaking stage $\delta(t)$$<$$\Gamma(t)$.}
     \end{figure*}
     \subsection{Knot-c: Model-fitting of inner kinematics (1995.5-2001.5)}
     In order to investigate the flux evolution associated with its 
     Doppler-boosting effect, the results of detailed studies on its 
     kinematics in the inner jet regions (1995.5-2001.5) are presented 
     in Figs.9--13.\\
      The traveled distance Z(t) of knot-c along the jet-axis is shown in 
     Fig.9 (left panel). The curves of parameters $\epsilon(t)$ and 
     $\psi(t)$ are presented in the right panel. Before 1996.40 
     (Z$\leq$13.8\,mas=106\,pc) $\epsilon$=$1.5^{\circ}$ and 
     $\psi$=--$42.8^{\circ}$ knot-c
     moved along the precessing common trajectory. After 1996.40 parameter 
     $\psi$ quickly increased to $\sim$--$20^{\circ}$ and knot-c started to
     move  along its own individual trajectory, deviating from the precessing
     common track. Such kind of transition from the common track pattern
     to the individual paths could involve some complex physical conditions
     (e.g., evolution of the kinetic and magnetic energy of the jet associated
      with the change in its current distribution, interaction between the
     jet and its surrounding environments, hydrodynamic and magnetohydrodynamic 
     instabilities (e.g., Hardee \cite{Hard87}, \cite{Har11};  Falke et al.
     \cite{Fa96}, Nakakura \& Meier \cite{Na04}, etc.).\\
      The model-fit of the trajectory $Z_n(X_n)$ during 1995.5--2001.5 is shown
     in Figure 10. Within coordinate $X_n$$\sim$0.46\,mas knot-c moved along
      the precessing common trajectory and beyond that knot-c started to move
      along its own individual track, deviating from the precessing common
      trajectory. In the figure the red and green curves are calculated for
      precession phases $\phi{\pm}$0.31\,rad ($\phi$=6.0\,rad for ejection at
      1995.54), indicating that the observational data-points are within the
      range defined by the two curves and the precession period is determined 
      with an uncertainty of $\sim$$\pm$5\% of the precession period (or 
      $\sim$$\pm$0.85\,yr.).\\
      The model-fits of core separation $r_n(t)$ (left panel), coordinates
      $X_n(t)$ and $Z_n(t)$ are  shown in Figure 11. Within 
      $r_n{\sim}$0.46\,mas knot-c moved along the precessing common trajectory
      and beyond that it started to move along its own individual track, 
      deviatitng from the common track. $r_n(t)$, $X_n(t)$ and $Z_n(t)$ are all
      well fitted during 1995.5--2001.5 (in the range of $r_n$ extending
      to 2.5\,mas).\\
      The model-derived apparent velocity ${\beta_{app}}(t)$, viewing angle
      $\theta(t)$, bulk Lorentz factor
      $\Gamma(t)$ and Doppler factor $\delta(t)$ are shown in Figure 12.
      ${\beta_{app}}(t)$, $\Gamma(t)$ and $\delta(t)$ have a bump stracture: 
      at 1996.15 $\delta$=$\delta_{max}$=30.2, $\Gamma$=29.7 and 
      $\beta_{app}$=29.6. At 1996.02  $\Gamma$=$\Gamma_{max}$=38.0 
      and $\beta_{app}$=${\beta_{app,max}}$=37.0.\footnote{In comparison, an
      average velocity $\beta_{app}$=22.9 was given in Britzen et al.
       \cite{Br17}.} $\theta(t)$ varied in the range of
     [$1.92^{\circ}$, $5.32^{\circ}$] during 1995.5--2001.0. It should be noted
      that during the peaking stage $\delta(t)$$<$$\Gamma(t)$.
    \subsection{Knot c: flux evolution and Doppler-boosting effect}
     \begin{figure*}
     \centering
     \includegraphics[width=5.6cm,angle=-90]{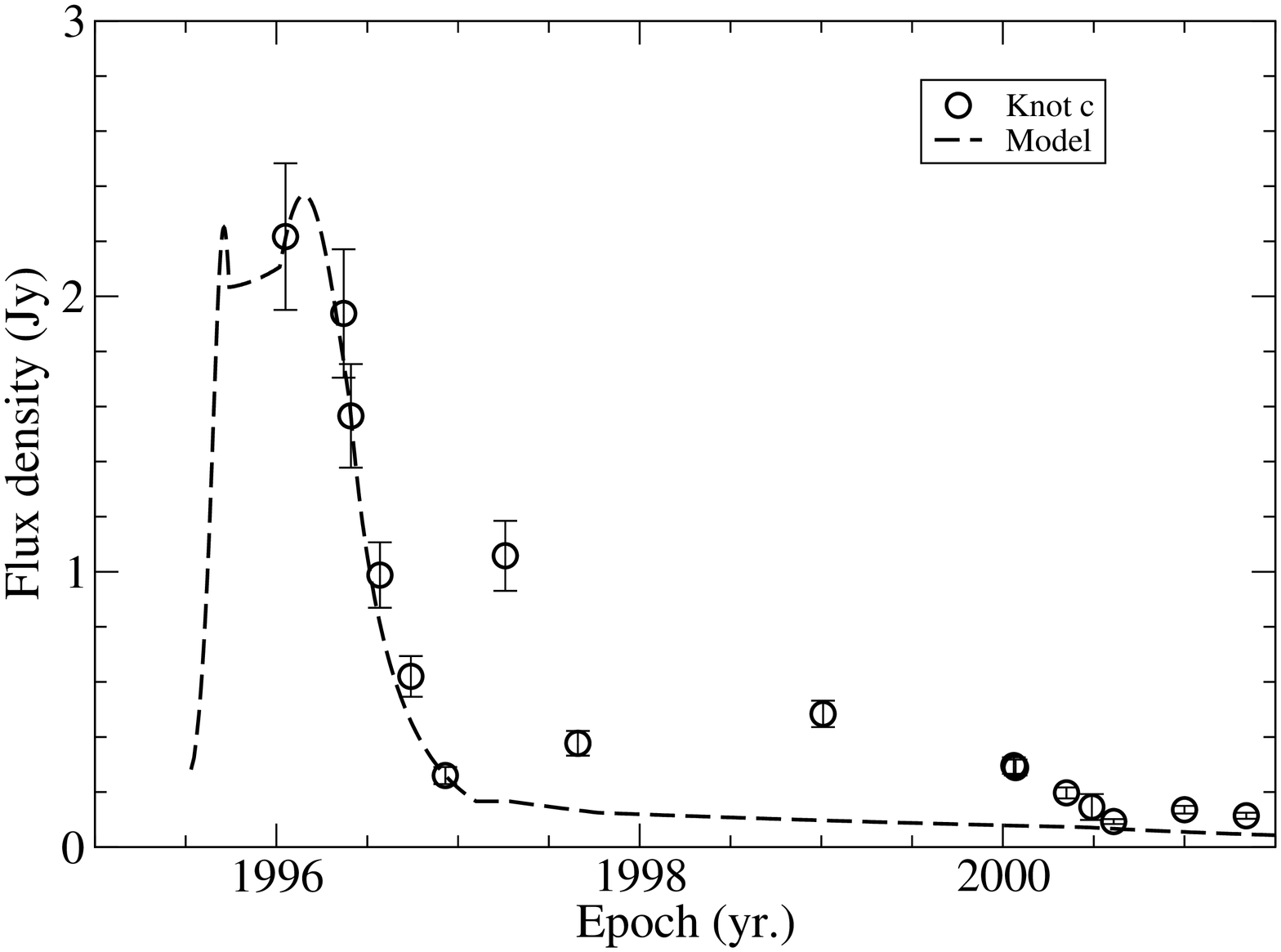}
     \includegraphics[width=5.6cm,angle=-90]{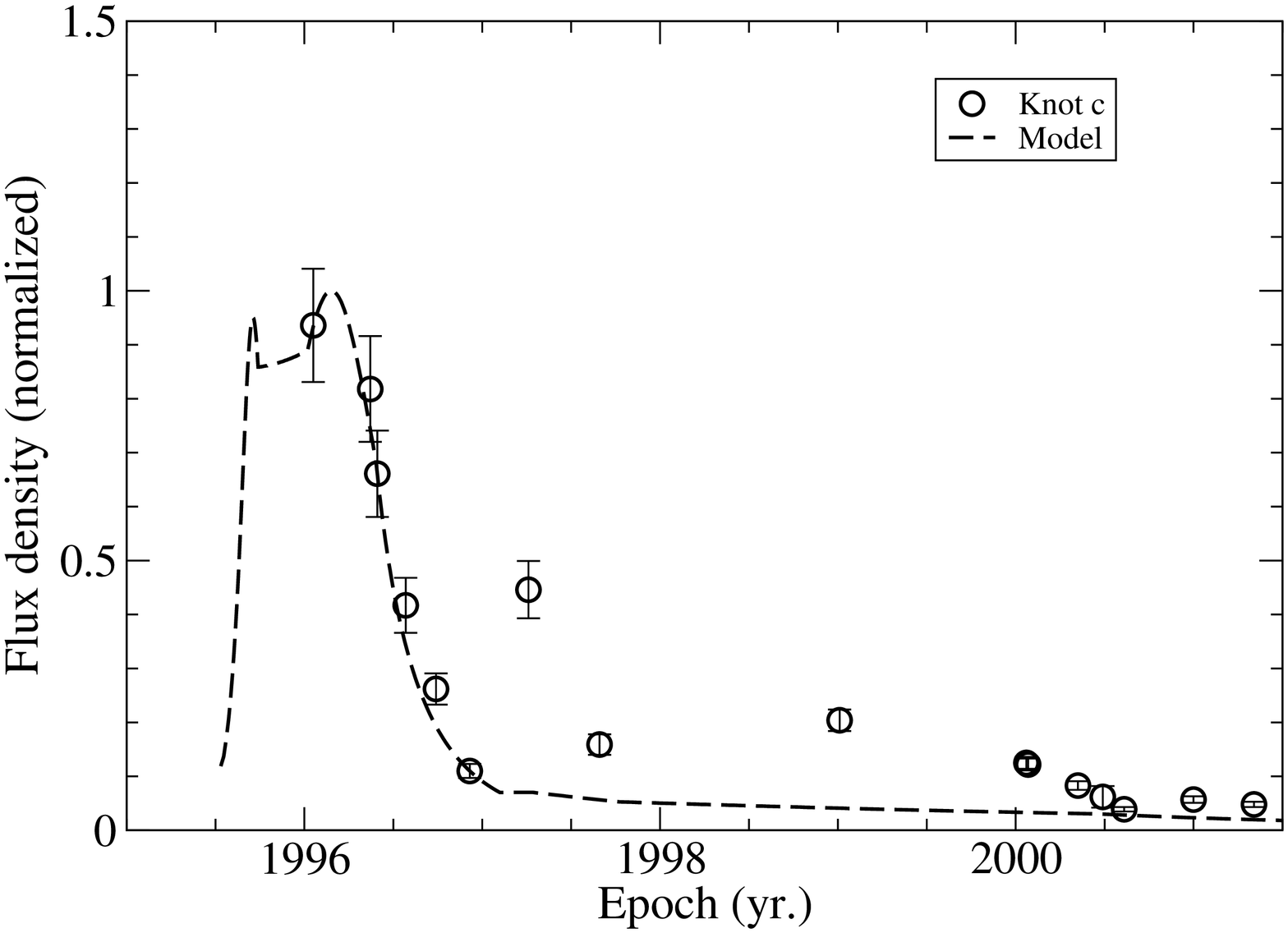}
     \caption{Knot-c. Left panel: the model-fit of the 15\,GHz light curve with
     the modeled peak flux density of 2.37\,Jy (at 1996.15) 
     and the intrinsic flux density
     of 15.6\,$\mu$Jy. Right panel: the light curve normalized by the modeled 
     peak flux density is very well fitted by the Doppler-boosting
      profile $[\delta(t)/\delta_{max}]^{3+\alpha}$ with an assumed value
      $\alpha$=0.5. The data-points at 1997.26, 1997.66 and 1997.01 deviating
     from the Doppler-boosting profile obviously could be due to variations 
    in its intrinsic flux density.}  
     \end{figure*}
     The model fit of the 15\,GHz light curve is shown in
     Figure 13. The modeled peak flux density is 2.37\,Jy (at 1996.15) and 
     its intrinsic flux density 15.6$\mu$Jy. The light curve
    normalized by the modeled peak flux density is well fitted by the
     Doppler-boosting profile $[\delta(t)/\delta_{max}]^{3+\alpha}$ (with
     an assumed value $\alpha$=0.5). The flux fluctuations on shorter 
     timescales (e.g. at 1997.26) which obviously deviate from the 
    Doppler-boosting  profile  might be induced by the intrinsic flux
    variations of knot-c itself.
   \begin{figure*}
  \centering
  \includegraphics[width=5.6cm,angle=-90]{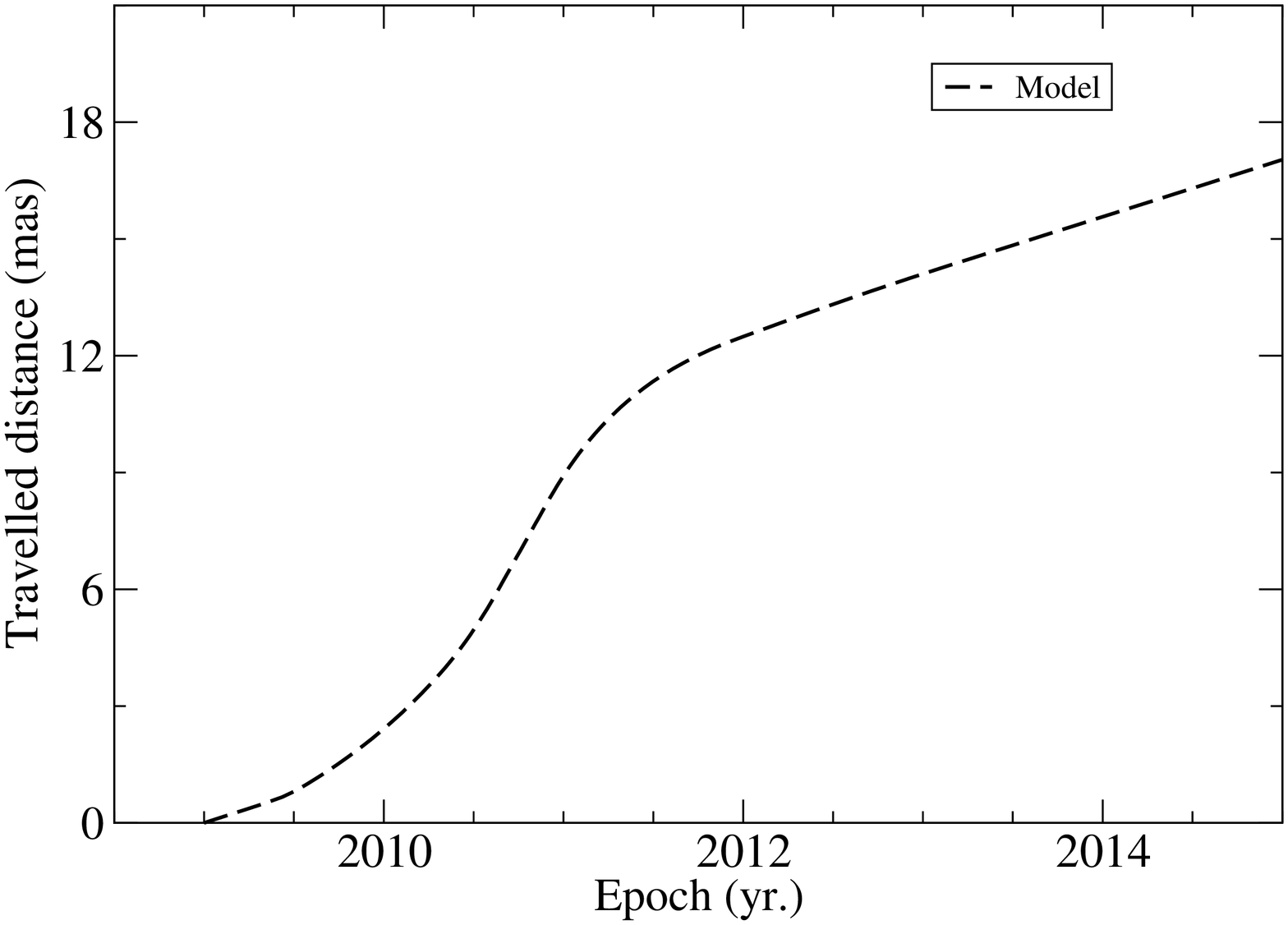}
  \includegraphics[width=5.6cm,angle=-90]{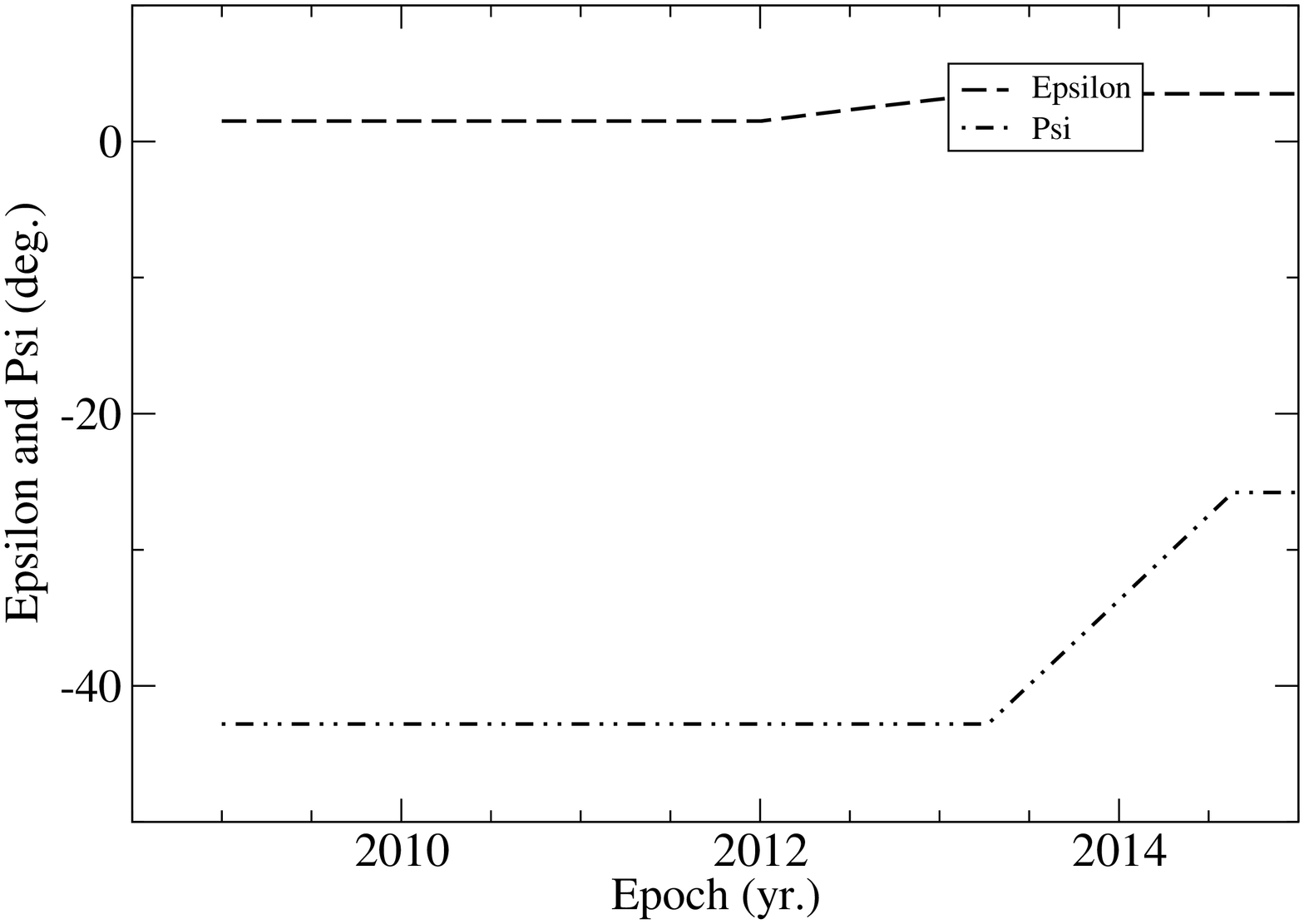}
  \caption{Knot-i. Left panel: the modeled traveled distance Z(t) along the 
   jet-axis. Right panel: the modeled curves for parameters $\epsilon(t)$
     and $\psi(t)$.  Before 2012.0 (or Z(t)
    $\leq$12.5\,mas=96.1\,pc) $\epsilon$=$1.5^{\circ}$ and 
   $\psi$=--$42.8^{\circ}$, knot-i moved along the precessing common trajectory.
   Beyond that distance its motion started to transfer to follow its own 
   individual track, deviating from the precessing common trajectory. }
  \end{figure*}
  \begin{figure*}
   \centering
   \includegraphics[width=6.5cm,angle=-90]{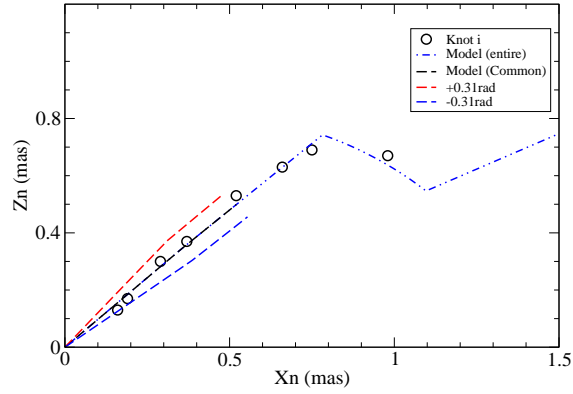}
   \caption{Knot-i: the model fit of the observed trajectory $Z_n(X_n)$.
    Within coordinate $X_n{\simeq}$0.39\,mas (Z$\leq$12.5\,mas=96.1\,pc),
     knot-i moved along the precessing
    common trajectory. Beyond that distance it started to follow its own 
   individual track. The red and green curves are calculated for precession
   phases $\phi{\pm}$0.31\,rad ($\phi$=1.0\,rad), demonstrating that the 
   observational data-points are within the area defined by the two curves 
   and the precession period is determined correctly with an uncertainty of
   $\sim$${\pm}$5\% of the precession period (or ${\sim}$$\pm$0.85\,yr.).}
   \end{figure*}
   \begin{figure*}
   \centering
   \includegraphics[width=5.6cm,angle=-90]{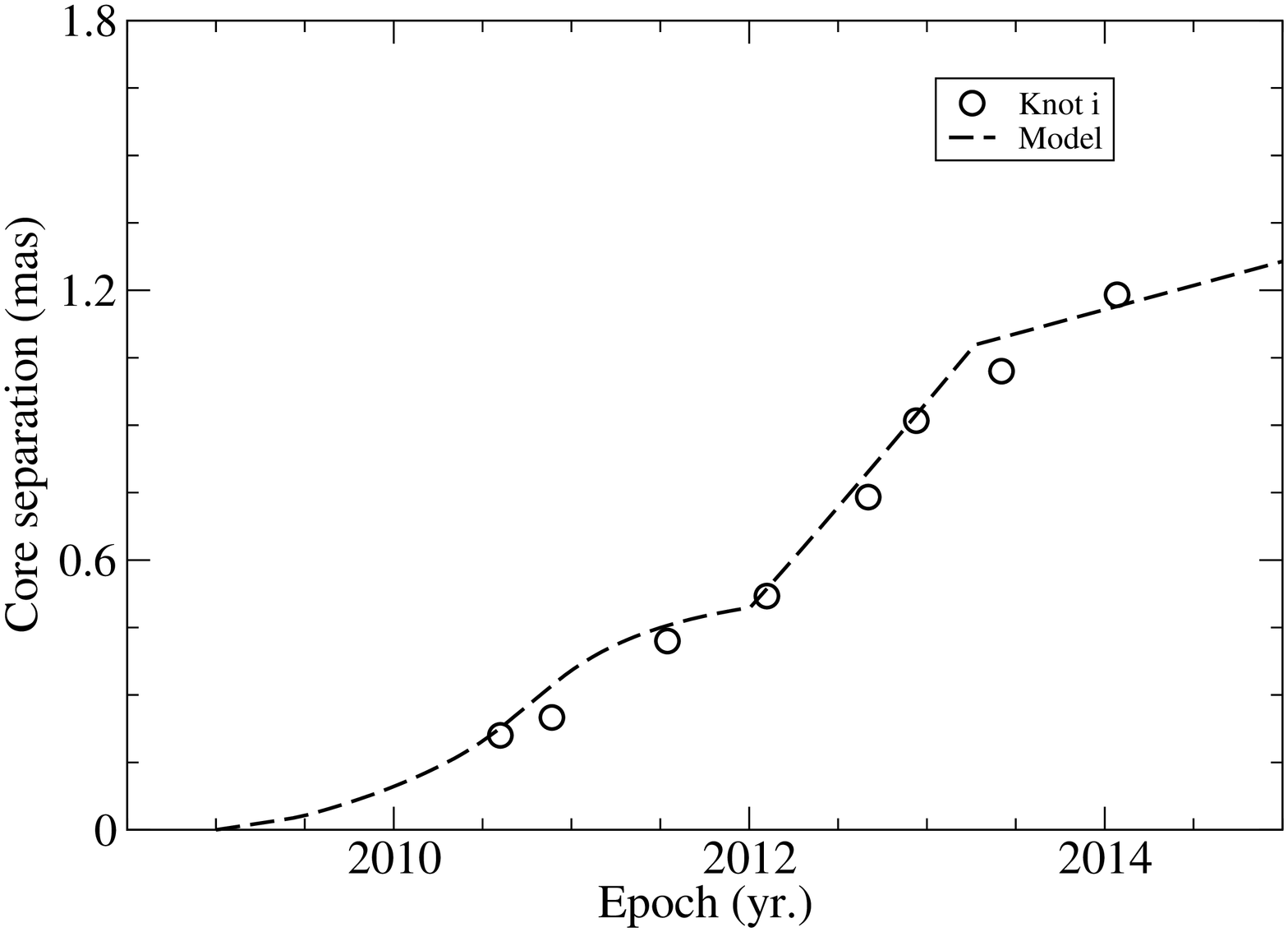}
   \includegraphics[width=5.6cm,angle=-90]{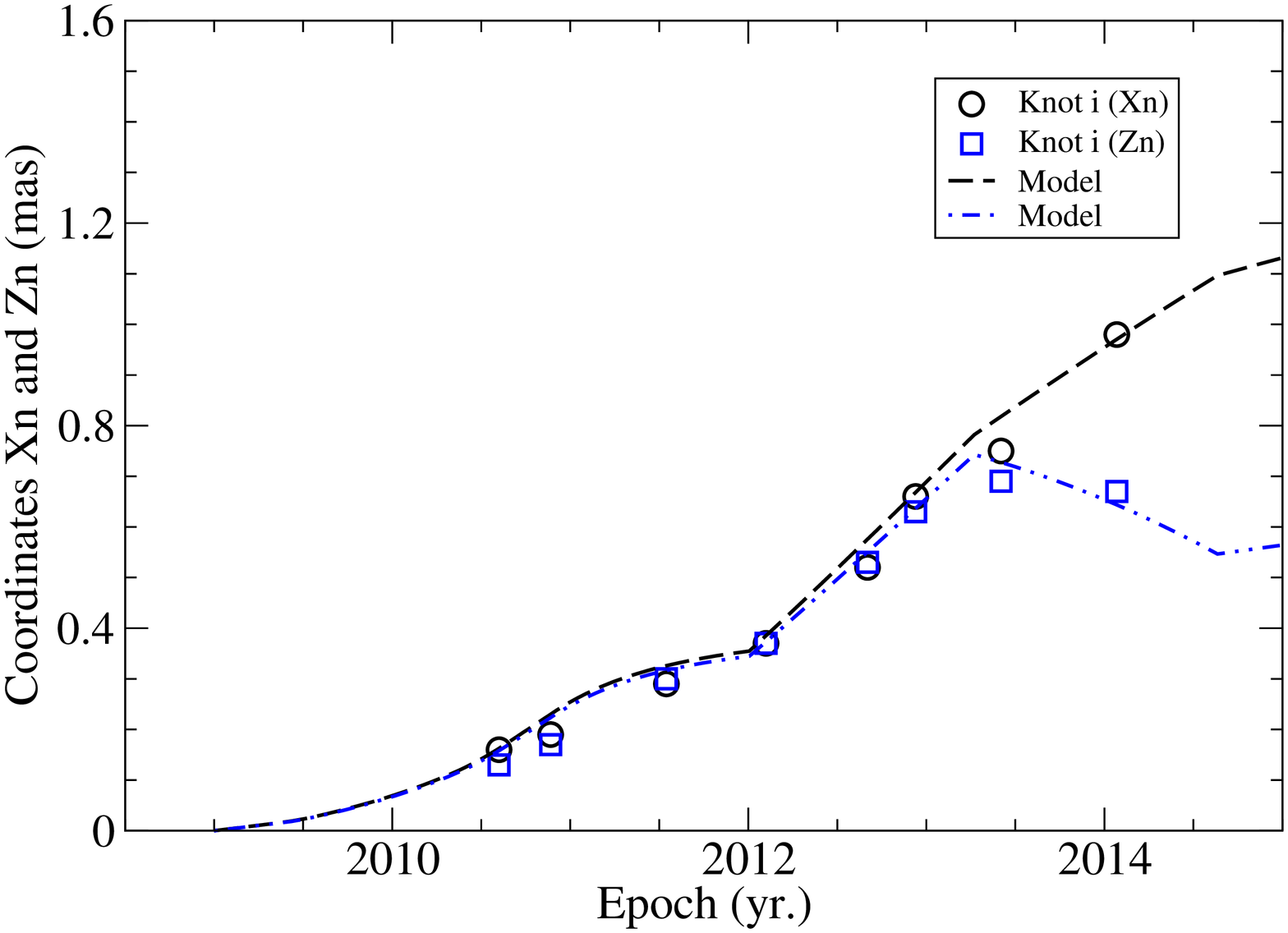}
   \caption{Knot-i. Left panel: the model fit of the core distance $r_n(t)$.
    Right panel: the model fits of the coordinates $X_n(t)$ and $Z_n(t)$.
    All the kinematic features are well
    fitted extending to $r_n\simeq$1.2\,mas (during 2010.5--2014.5).}
   \end{figure*}
   \begin{figure*}
   \centering
   \includegraphics[width=5.6cm,angle=-90]{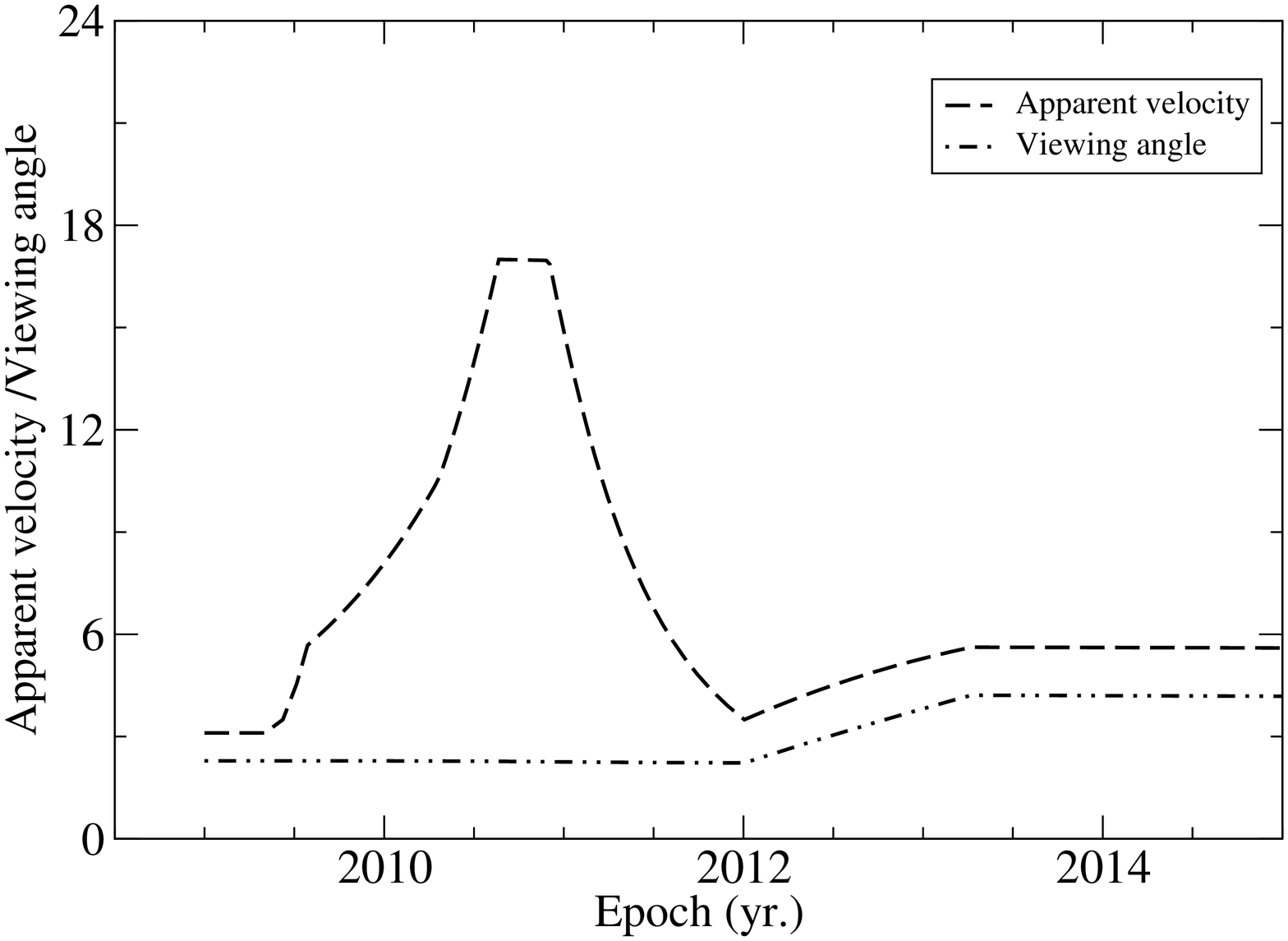}
   \includegraphics[width=5.6cm,angle=-90]{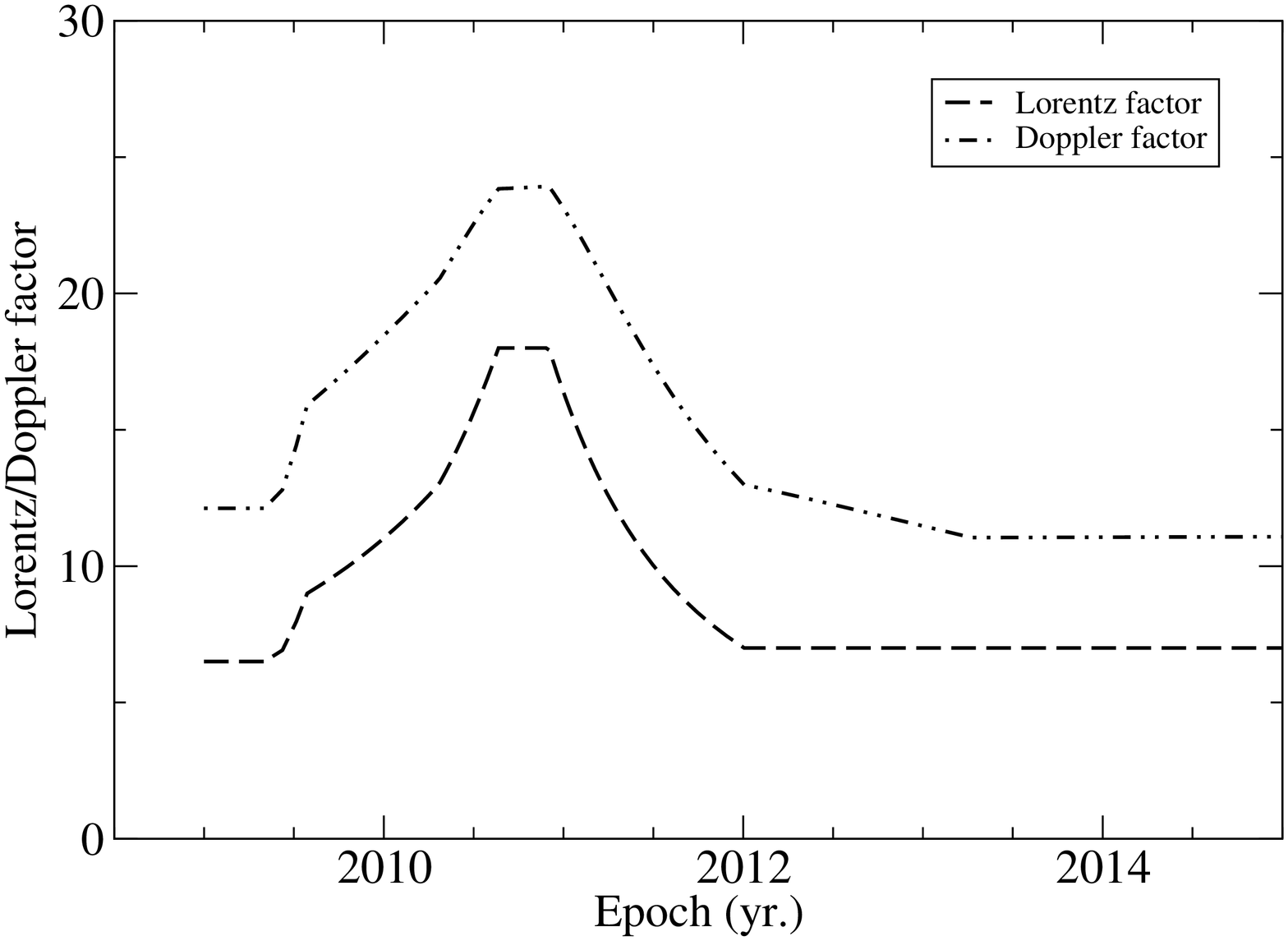}
   \caption{Knot-i. Left panel: the model-derived apparent velocity 
   $\beta_{app}(t)$ and 
    viewing angle $\theta(t)$. Right panel: the  model-derived bulk
    Lorentz factor $\Gamma(t)$ and Doppler factor $\delta(t)$. $\beta_{app}(t)$,
    $\Gamma(t)$ and $\delta(t)$ all have a bump structure during 2009.5--2012.0,
   which was  associated with a major burst (see Figure 18 below). At 2010.90
   $\delta$= $\delta_{max}$=24.02 and $\Gamma$=$\Gamma_{max}$=13.80. At 2010.64 
    $\beta_{app}$=$\beta_{app,max}$=16.96. $\theta(t)$ varied in the range 
   of  [$2.27^{\circ}$, $4.18^{\circ}$] during 2009.5--2014.0. }
   \end{figure*}
   \begin{figure*}
   \centering
   \includegraphics[width=5.6cm,angle=-90]{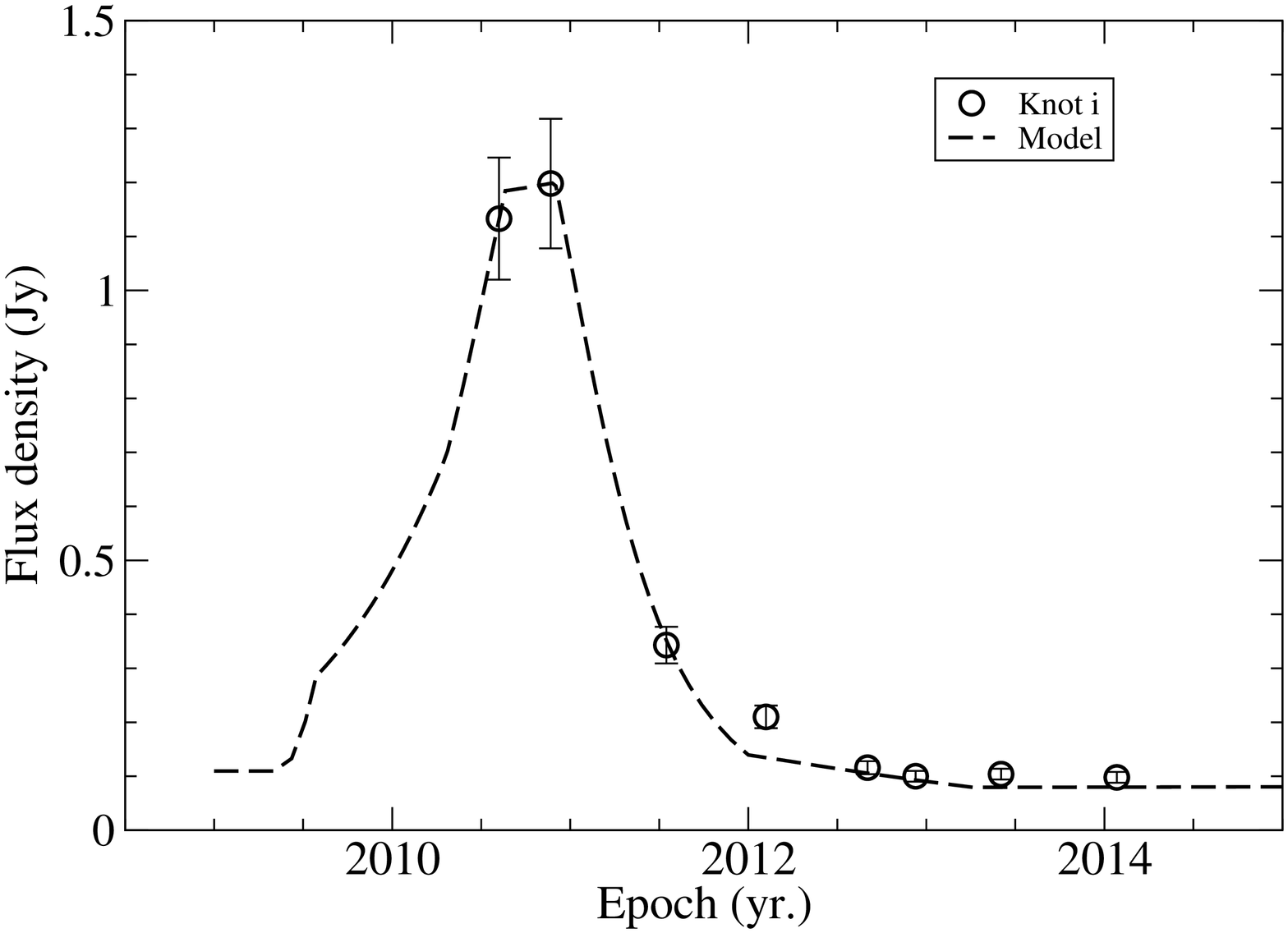}
   \includegraphics[width=5.6cm,angle=-90]{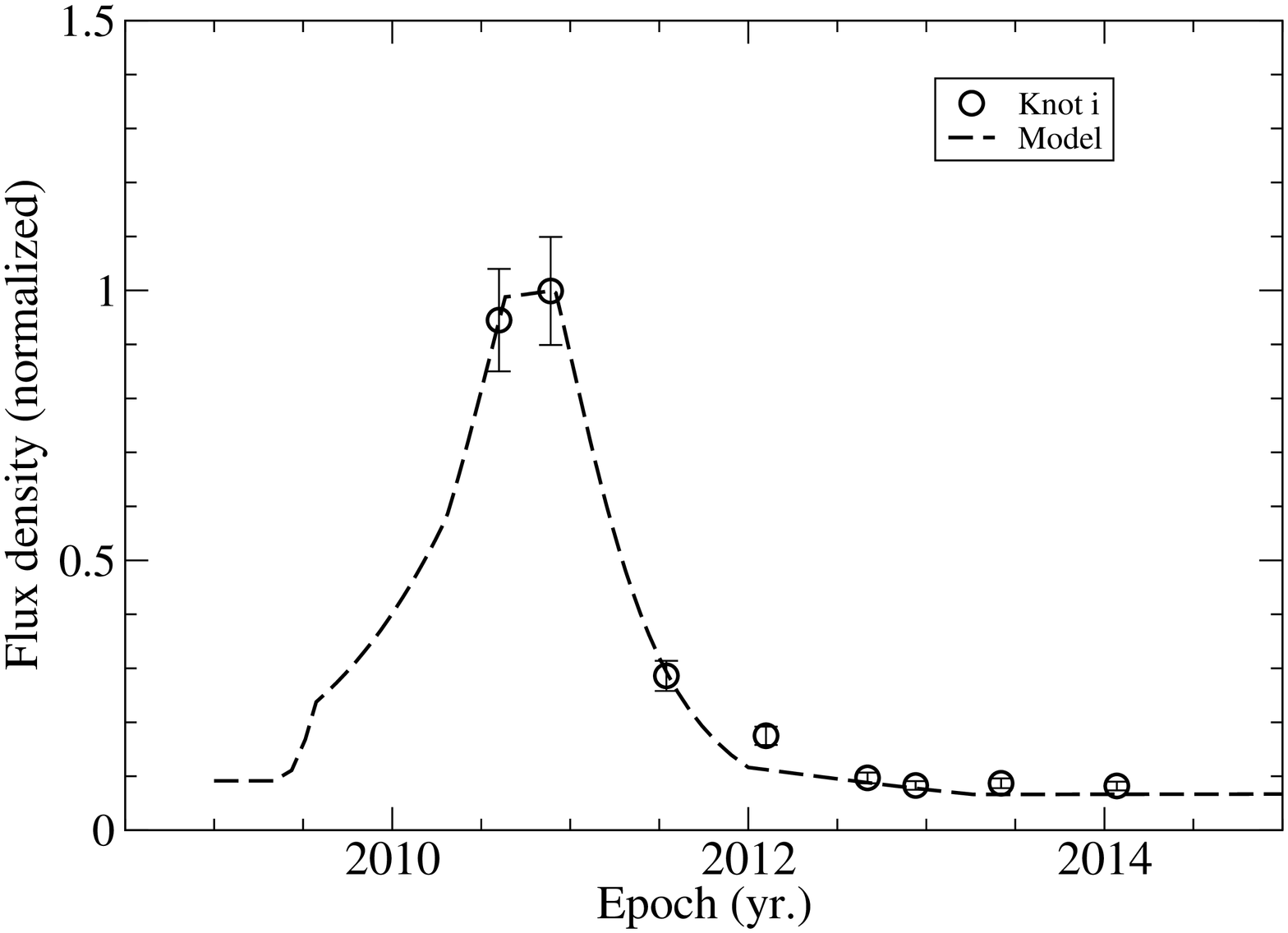}
   \caption{Knot-i. Left panel: the 15GHz light curve is well fitted  with
   its modeled peak flux density of 1.20\,Jy at 2010.90 and intrinsic flux 
   density of 17.7$\mu$\,Jy. Right panel: the light curve
    normalized by the modeled peak flux density is fitted very well by 
   the Doppler-boosting profile $[\delta(t)/\delta_{max}]^{3+\alpha}$. }
   \end{figure*}
   \section{Knot i: Model-fitting results}
     The model-fitting results for knot-i are shown in Figures 14--18.
     Its ejection time $t_0$=2009.0 and the corresponding precession
     phase $\phi$=1.0\,rad. 
    \subsection{Knot-i: model-fit of kinematics}
    The modeled traveled distance Z(t) along the jet-axis and the modeled 
     curves for parameters $\epsilon(t)$ and $\psi(t)$ are shown in Figure 14.
     Before 2012.00 (Z$\leq$12.5\,mas=96.1\,pc) $\epsilon$=$1.5^{\circ}$ and 
     $\psi$=--$42.8^{\circ}$ knot-i moved along the precessing common 
     trajectory, while after that $\epsilon$ increased knot-i started to move 
     along its own individual track, deviating from the common precessing
     track ($\psi$ started to increase quickly after 2013.38).\\
     The model fit of its projected trajectory $Z_n(X_n)$ is shown in Figure 
    15. Within core separation $r_n$$\sim$0.49\,mas knot-i moved along the
     common precessing track. Beyond that knot-i started to move along its
    own individual track, deviating from the common track. The red and green
    curves in the figure are calculated for precession phase 
    $\phi{\pm}$0.31\,rad ($\phi$=1.0\,mas), indicating that the observational 
    data-points are within the area defined by the two curves and the
    precession period is determined with an uncertainty of $\sim$${\pm}$5\%.\\
    Model fits of core separation $r_n(t)$ (left panel), coordinates $X_n(t)$
    and $Z_n(t)$ are shown in Figure 16. Before 2012.00 ($r_n$$\leq$0.49\,mas,
    $X_n$$\leq$0.39\,mas) knot-c moved along the precessing common track,
    while after that knot-c stared to move along its own individual 
   trajectory, deviating
    from the common precessing track. It can be seen that $r_n(t)$, $X_n(t)$ 
    and $Z_n(t)$ are all well model-fitted during 2010.5--2014.0. \\
    The model-derived apparent velocity $\beta_{app}(t)$, viewing angle
    $\theta(t)$, bulk Lorentz factor
     $\Gamma(t)$ and Doppler factor $\delta(t)$ are shown in figure 17.
    $\beta_{app}$, $\Gamma$ and $\delta$ all have a bump structure coincident 
    with the radio burst (see Figure 18).
    At 2010.90 $\delta$=$\delta_{max}$=24.02, $\Gamma$=$\Gamma_{max}$=18.0. 
     At 2010.64 $\beta_{app}$=$\beta_{app,max}$=16.94. $\theta(t)$ varied 
    in the range of [$2.27^{\circ}$ (2009.0), $4.18^{\circ}$ (2014.0)].
    \subsection{Knot-i: flux evolution and Doppler-boosting effect}
     The model fit of the 15GHz light curve is shown in Fig.18 (left panel)
    with its modeled peak flux density 1.20\,Jy (at 2010.90) and the intrinsic
     flux density 17.7$\mu$Jy. The light curve  normalized by the modeled peak 
    flux density was well fitted by its Doppler-boosting profile 
    $[\delta(t)/\delta_{max}]^{3+\alpha}$ (right panel).
    \section{Knot k: Model-fitting results}
     The model fitting results of the kinematics and light curve for knot-k 
    are shown in Figures 19--23. Its ejection time $t_0$=2010.88 and the 
    corresponding precession phase $\phi$=0.30\,rad.
    \subsection{Model-fitting of kinematics}
     The modeled traveled distance Z(t) along the jet-axis and the modeled 
    curves for parameters $\epsilon(t)$ and $\psi(t)$ are shown in Figure 19.
    Before 2012.97 (or Z$\leq$7.2\,mas=55.4\,pc) $\epsilon$=$1.5^{\circ}$ and 
   $\psi$=--$42.8^{\circ}$, knot-k moved along the precessing common track 
   and after that $\epsilon$ decreased slightly knot-k started to move along 
   its own individual track, slightly deviating from the precessing common
    trajectory.\\
     The model-fit of the projected trajectory is shown in Figure 20.
    Within $X_n$=0.24\,mas (or before 2012.97) knot-k moved along 
    the precessing common track. Beyond that knot-k started to move along its
    own individual track, slightly deviating from the common 
    precessing track. The red and green curves
   in the Figure are calculated for precession phases $\phi{\pm}$0.31\,rad 
   ($\phi$=0.30\,rad), indicating that the observational data-points are within
   the area defined by the two curves and the precession period is determined
   with an uncertainty of $\sim$$\pm$5\% of the period 
    ($\sim$$\pm$0.85 years).\\
    The model-fits of the core separation $r_n(t)$,  
    coordinates $X_n(t)$ and $Z_n(t)$ are shown in Figure 21. Within 
   $r_n$=0.26\,mas (or $X_n$$\leq$0.24\,mas; before 2012.97) knot-k moved along 
    the precessing common track, while
     beyond that it started to move along its own individual track, 
   deviating from the common track.\\
      The model-derived  apparent velocity $\beta_{app}(t)$, viewing 
    angle $\theta(t)$, bulk Lorentz factor $\Gamma(t)$ and Doppler factor 
    $\delta(t)$ are shown in Figure 22. $\beta_{app}$,
     $\Gamma$ and $\delta$ all 
    have a bump structure: at 2013.00 $\delta$=$\delta_{max}$=24.50,
    $\Gamma$=16.06, $\beta_{app}$=13.62 and $\theta$=$1.99^{\circ}$.
     But $\Gamma_{max}$=16.50 
    (during 2012.81--2012.97) and $\beta_{app,max}$=14.4 (at 2012.81).
    Viewing angle $\theta(t)$ varied in the range of 
    [$2.07^{\circ}$, $1.53^{\circ}$] during 2011.4--2014.0.
    \\
    \subsection{Knot k: flux evolution and Doppler-boosting effect}
    The model-fit of the 15\,GHz light curve is shown in Fig.22 (left panel)
    with its modeled peak flux density 1.26\,Jy (at 2013.00) and intrinsic flux
    density $S_{int}$=17.4\,$\mu$Jy. The light curve  normalized by the modeled
     peak flux density was well fitted by its  Doppler-boosting profile 
     $[\delta(t)/\delta_{max}]^{3+\alpha}$ with a presumed value $\alpha$=0.5
     (right panel). The data-point at 2014.07, obviously deviating from
    the profile, might be due to the increasing of its intrinsic flux density.
     \begin{figure*}
     \centering
     \includegraphics[width=5.6cm,angle=-90]{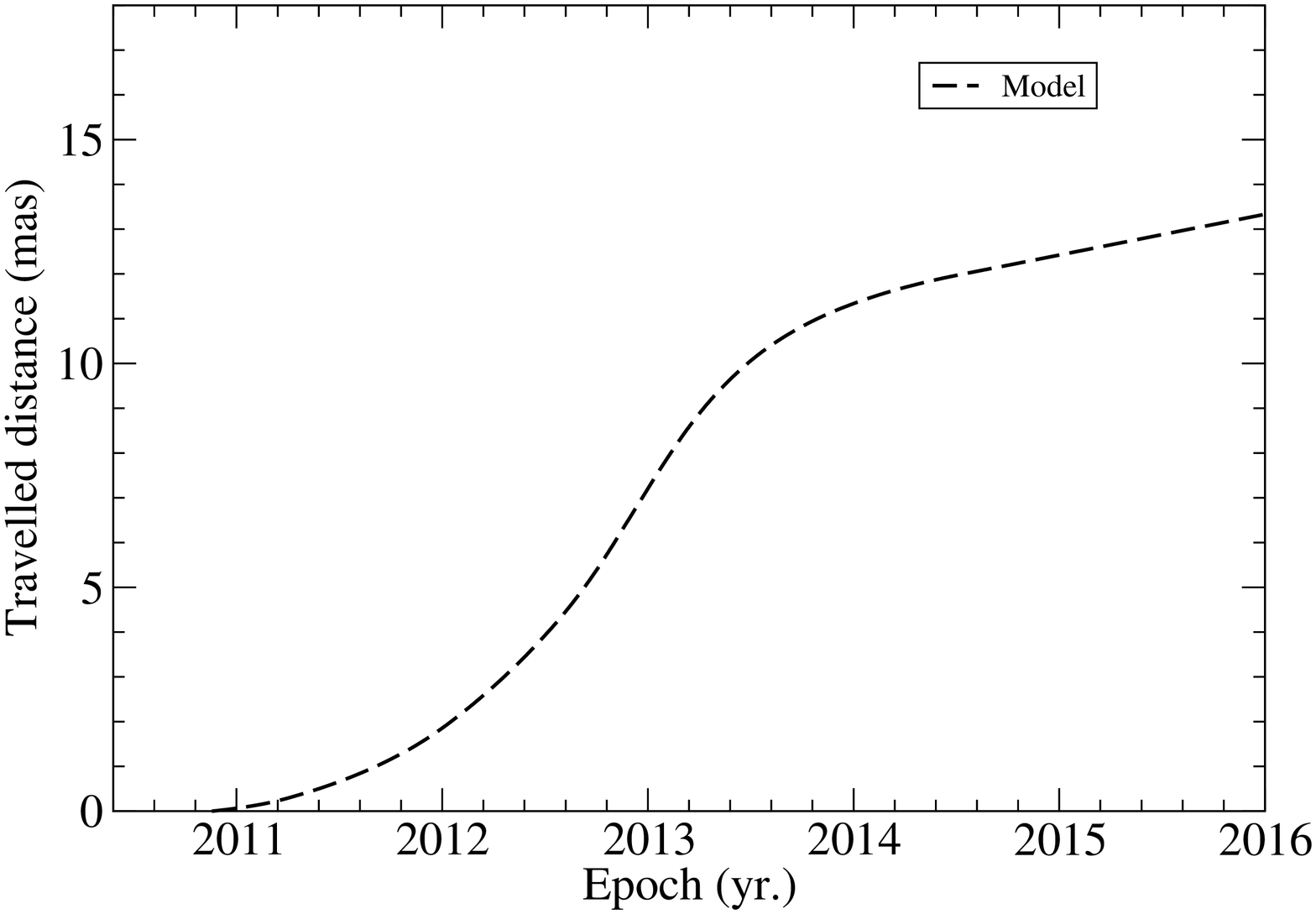}
     \includegraphics[width=5.6cm,angle=-90]{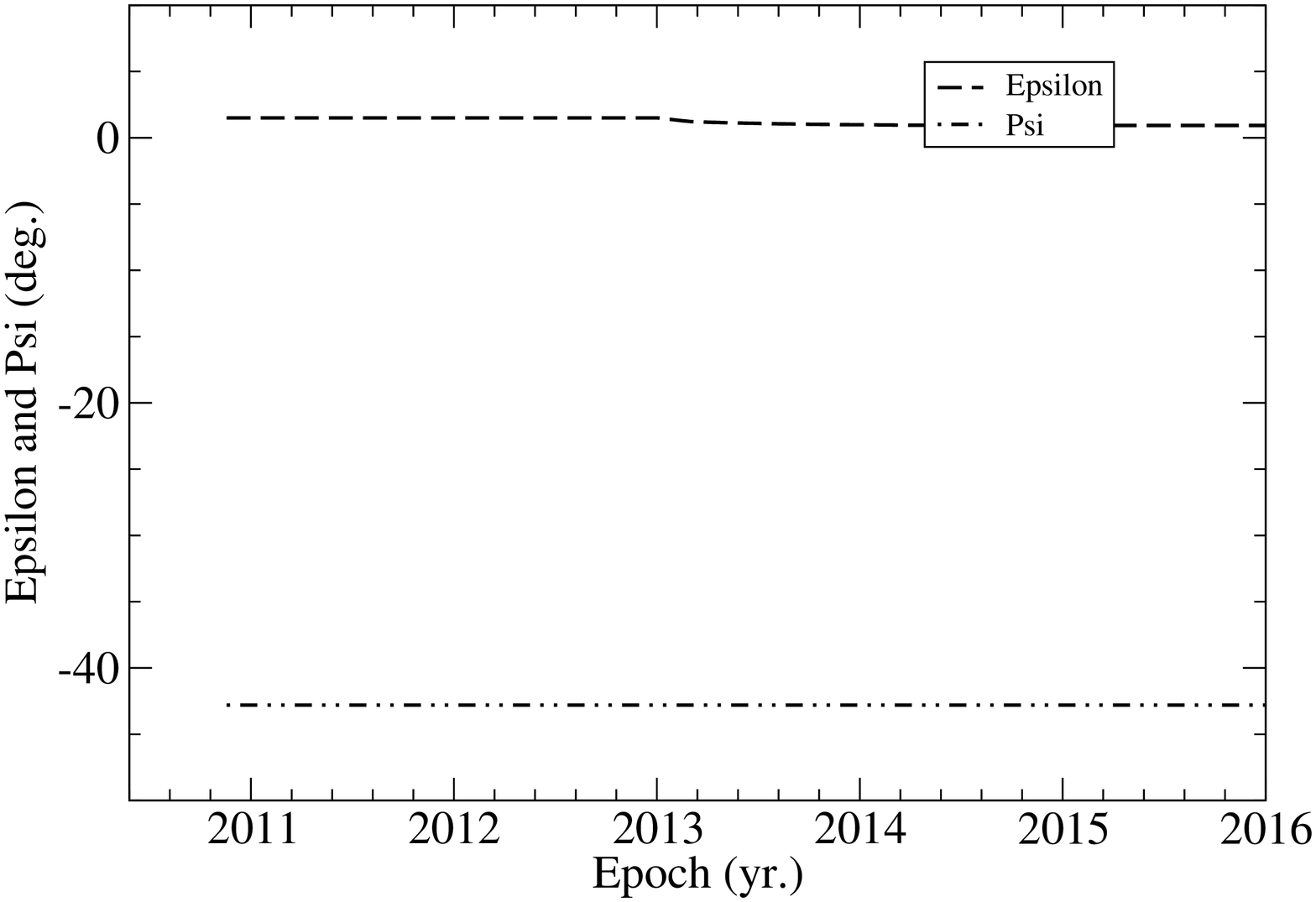}
     \caption{Knot-k. Left panel: the model-derived traveled distance Z(t).
    Right panel: the modeled curves of parameter $\epsilon(t)$ and $\psi(t)$.
     Before 2012.97 (corresponding to   
     traveled distance Z$\leq$7.2\,mas=55.4\,pc) $\epsilon$=$1.5^{\circ}$ and
     $\psi$=--$42.8^{\circ}$, knot-k moved along the precessing common 
     trajectory. After that epoch $\epsilon$ slightly decreased with time and 
     knot-k  started to move along its own individual trajectory, slighly
    deviating from the precessing common track.}
     \end{figure*}
     \begin{figure*}
     \centering
     \includegraphics[width=6.5cm,angle=-90]{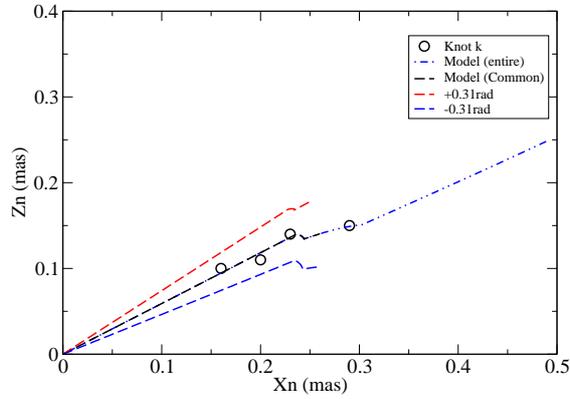}
    \caption{Knot-k: the model fit of the observed trajectory $Z_n(X_n)$.
    Within $X_n{\leq}$0.24\,mas (Z$\leq$7.2\,mas=55.4\,pc) 
    knot-k moved along the preceessing common 
    track. After that knot-k started to move along its own individual 
    trajectory, deviating from the precessing common track. The red and green
    curves were calculated for the precession phases $\phi{\pm}$0.31\,rad 
   ($\phi$=0.30\,rad), indicating that the observational data-points are within
    the area defined by the two curves and the precessing period is derived 
    correctly with an uncertainty of $\sim$$\pm$5\% of the period 
    (or ${\sim}$$\pm$0.85\,yr.).}
    \end{figure*}
     \begin{figure*}
     \centering
     \includegraphics[width=5.6cm,angle=-90]{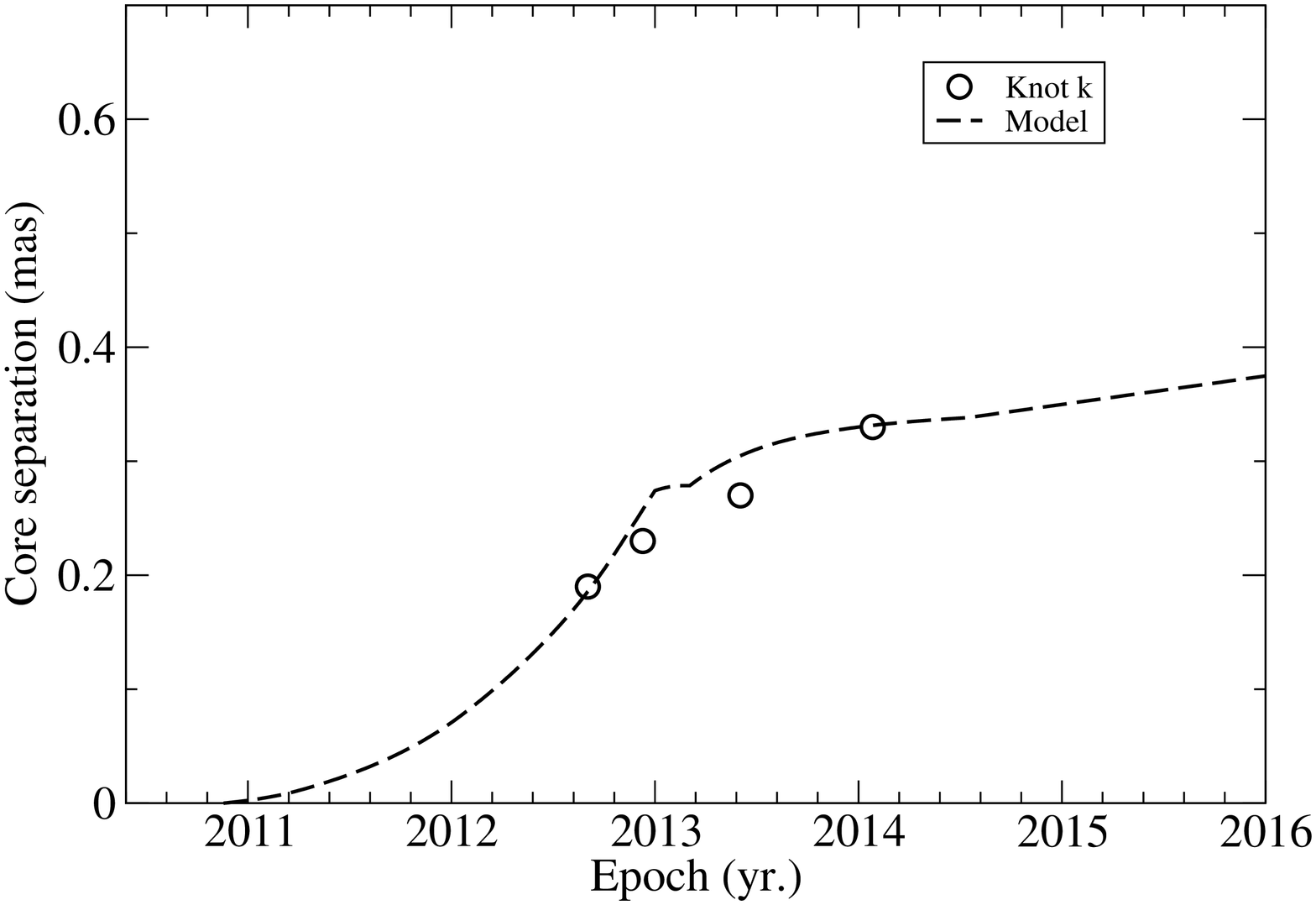}
     \includegraphics[width=5.6cm,angle=-90]{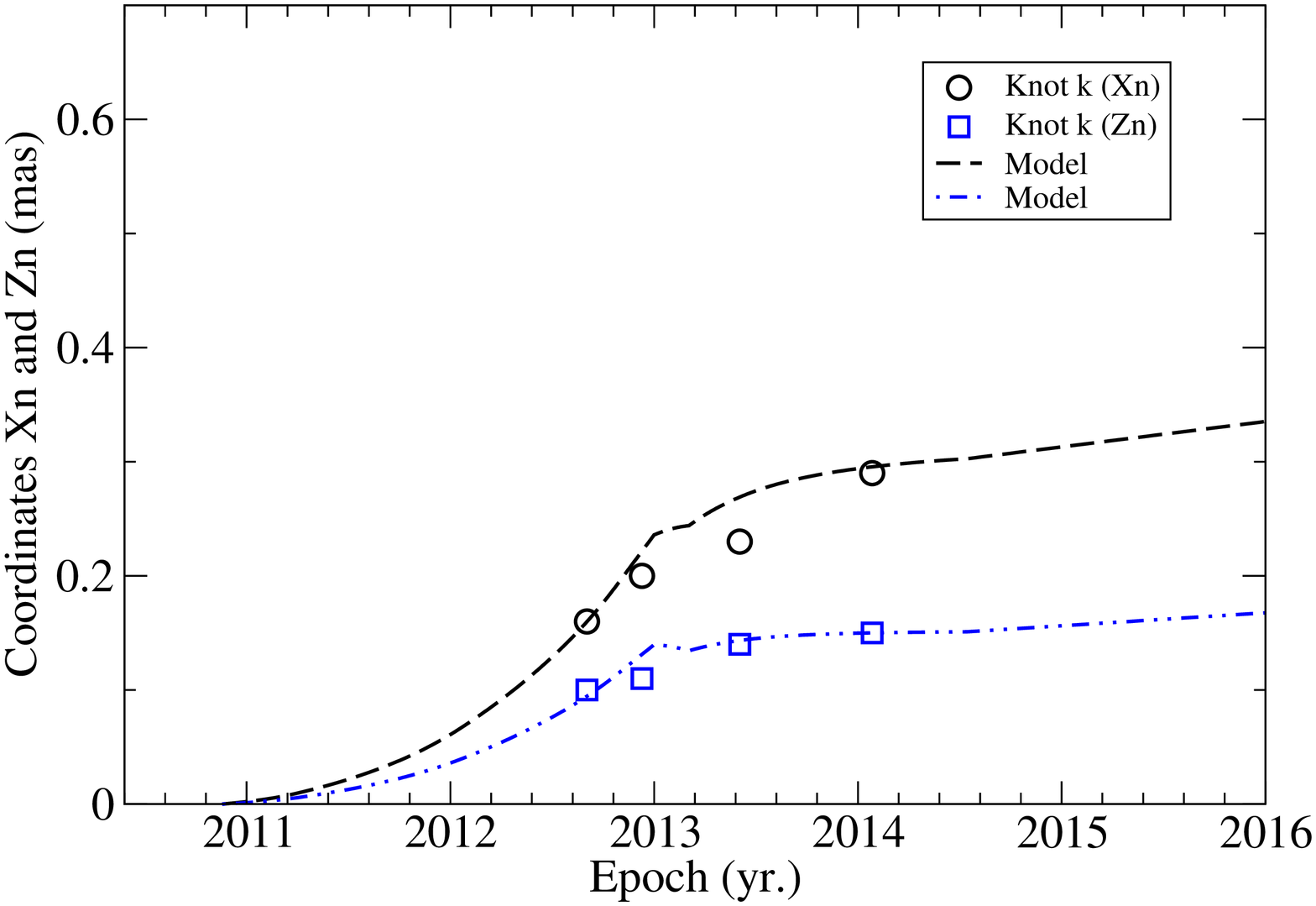}
     \caption{Knot-k. Left panel: the model fit of the core separation
     $r_n(t)$. Right panel:
     the model fits of the coordinates $X_n(t)$ and $Z_n(t)$. Before 2012.97
      (or $r_n{\leq}0.26$\,mas, $X_n{\leq}$0.24\,mas) $\epsilon$=$1.5^{\circ}$
      and $\psi$=--$42.8^{\circ}$, knot-k moved along the preceessing common 
     track. After that knot-k started to move along its own individual 
     trajectory, deviating from the precessing common track.}
     \end{figure*}
      \begin{figure*}
     \centering
     \includegraphics[width=5.6cm,angle=-90]{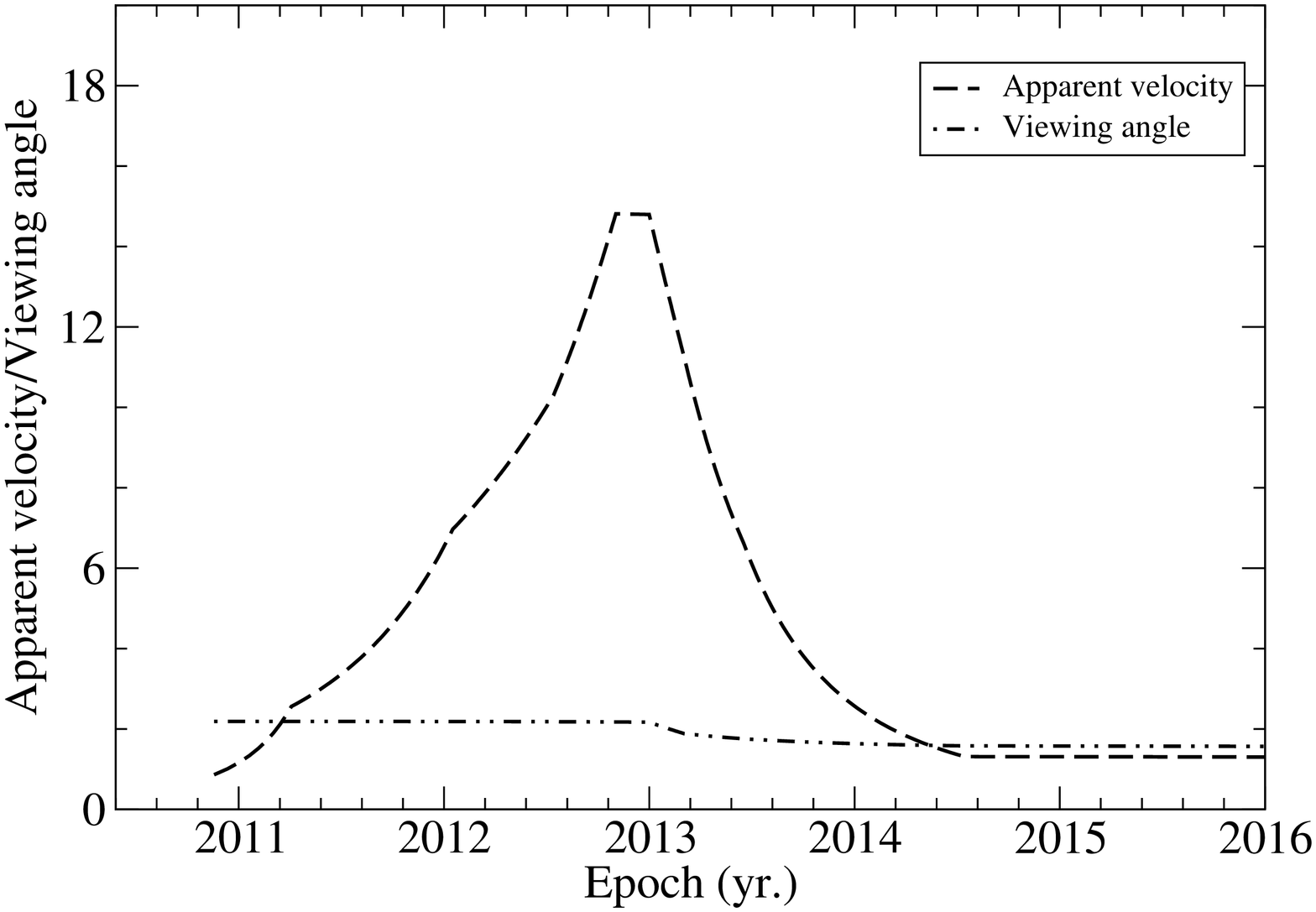}
     \includegraphics[width=5.6cm,angle=-90]{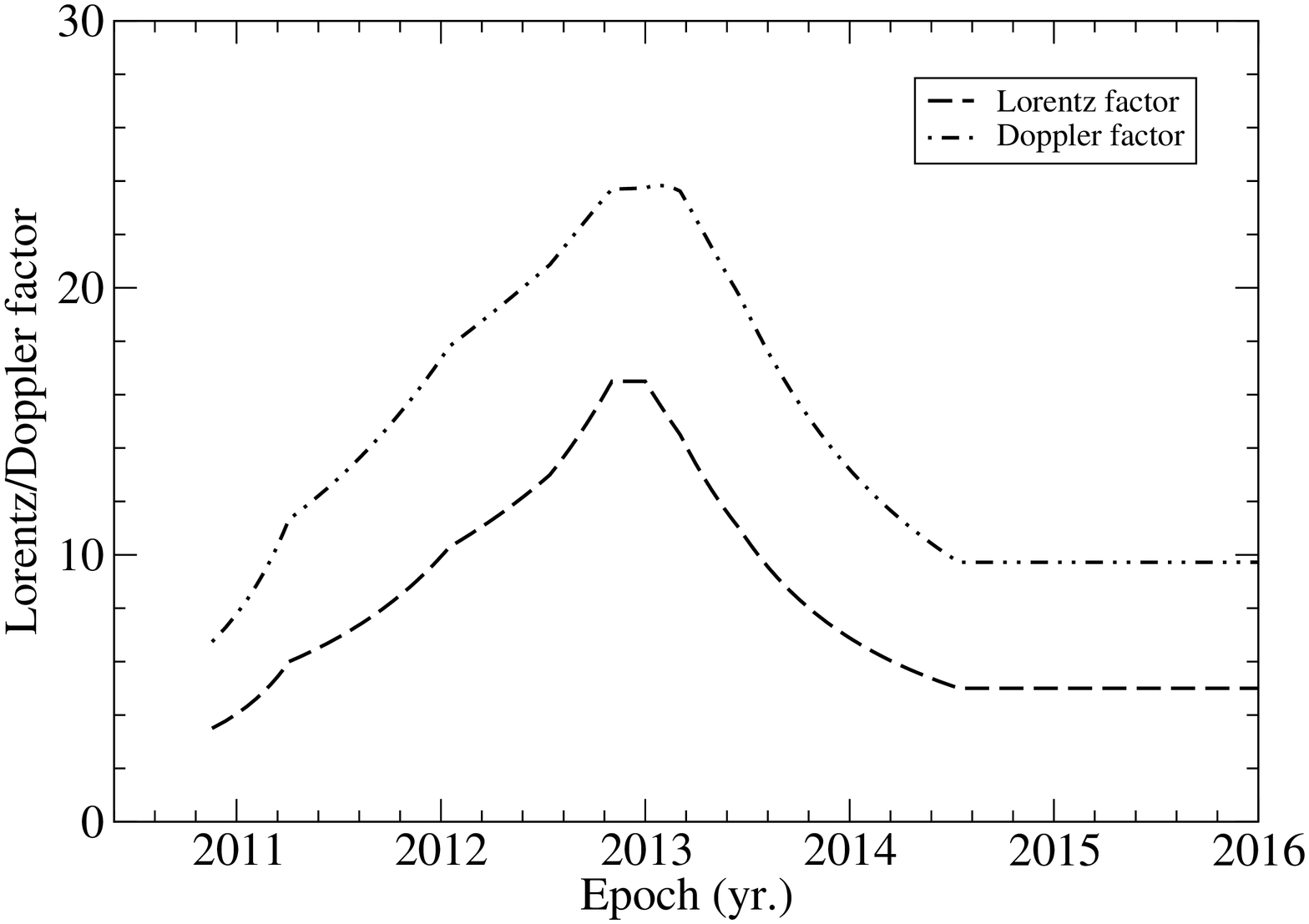}
     \caption{Knot-k. Left panel: the model-drived apparent velocity $\beta_{app}(t)$ and
     viewing angle $\theta(t)$. Right panel: the  model-drived bulk Lorentz 
     factor $\Gamma(t)$ and Doppler factor $\delta(t)$. At 2013.00 $\delta$=
     $\delta_{max}$=24.50, $\Gamma$=16.1 and $\beta_{app}$=13.6, while 
      $\Gamma$=$\Gamma_{max}$=16.5 during  2012.81--2012.97 and   
    $\beta_{app}$=$\beta_{app,max}$=14.4 at 2012.81. $\theta(t)$ varied in
     the range of [$2.07^{\circ}$, $1.53^{\circ}$] during 2011.4--2014.0.}
     \end{figure*}
     \begin{figure*}
     \centering
     \includegraphics[width=5.6cm,angle=-90]{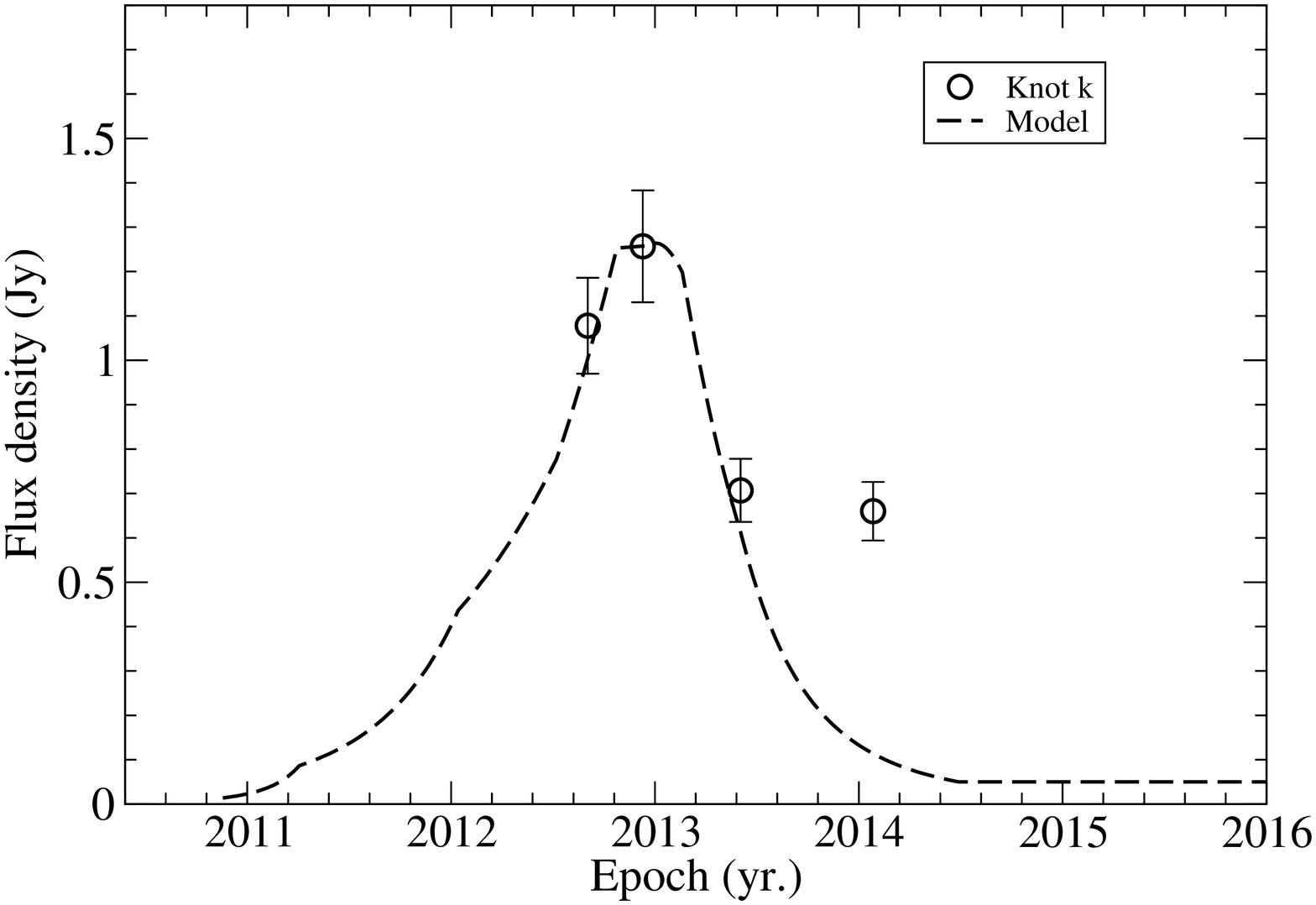}
     \includegraphics[width=5.6cm,angle=-90]{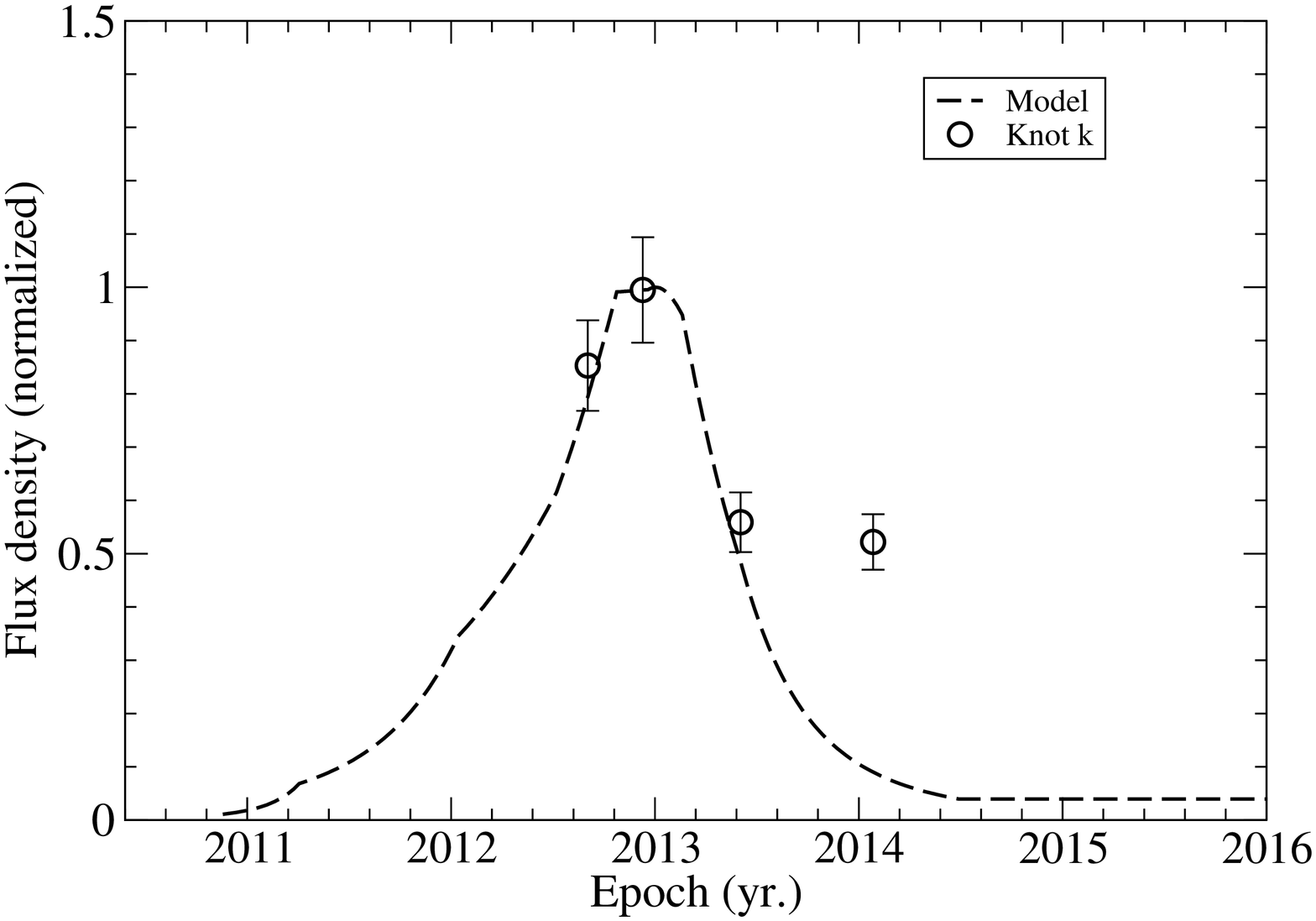}
     \caption{Knot-k. Left panel: the model fit of the 15\,GHz light curve
     with its modeled peak flux density of 1.26\,Jy (at 2013.0) 
     and intrinsic flux density
      of 17.4\,$\mu$Jy.  Right panel: the light curve normalized by the
     modeled peak flux density is well fitted by the Doppler-boosting profile
      $[\delta(t)/\delta_{max}]^{3+\alpha}$ with an assumed $\alpha$=0.5.
      The data-point at 2014.07 deviating from the profile largely might be
      due to a variation in its intrinsic flux density.}
     \end{figure*}
    \section{Summary and conclusion}
    Based on our precessing jet-nozzle secnario (Qian et al. \cite{Qi17},
    \cite{Qi21})
    we have successfully model-fitted the kinematics observed at 15\,GHz 
    on pc-scales for the three superluminal components (knot-c, -i and -k)
    in QSO B1308+326 and interpreted their light curves. We briefly summarize
    the results as follows:\\
    (1) Superluminal components were ejected from the precessing jet-nozzle
      with a precession period of $\sim$16.9$\pm$0.85\,years. \\
    (2) The superluminal knots moved along the precessing common tracks 
    corresponding to their  precession phases (or ejection times) 
    in the innermost jet regions (core separation $r_n\leq$0.30--0.5\,mas),
     while in the outer jet-regions
    they started to deviate from the precessing common tracks and moved along
     their own individual trajectories.\\
    (2) the periodic position angle swing observed for the superluminal 
    components can be  well explained in terms of our precessing nozzle
    scenario.\\
    (3) The observed kinematic features (trajectory $Z_n(X_n)$, 
     core separation ($r_n(t)$), coordinates $X_n(t)$ and $Z_n(t)$ and
     apparent velocity $\beta_{app}(t)$) were consistently  well model-fitted.\\
    (4) The bulk Lorentz factor $\Gamma(t)$, viewing angle $\theta(t)$ and
     Doppler factor $\delta(t)$ for the superluminal components  were 
     properly derived.\\
    (5) The 15\,GHz light curves of the superluminal components 
     can be well interpreted in terms of their Doppler-boosting effect.\\
      Our precessing jet-nozzle scenario may be described by such a
     physically feasible conception:\\
       In the nucleus of B1308+328 there is a energy-engine consisting of a
     rotating (Kerr) black-hole and a tilted magnetized accretion-disk around
     the hole. Due to the electromagnetic effects induced from the spin 
     of the black hole and the rotation 
     of the disk with its magnetosphere, a mini-jet (or beam) is formed with 
     a nozzle steadily ejecting magnetized plasma and superluminal plasmoids
     along helical tracks around the axis of the disk. Moreover, 
      due to the frame-dragging effect the gravitational torque of the rotating
     black hole will cause a global precession of the accretion disk with its 
     mini-jet. Thus the precession of the mini-jet would naturally form the
     precessing common trajectory suggested in our scenario, producing the 
     observed periodic position angle swing of superlumnal components
     and regular distribution of their inner trajectories. Obviously, 
     in our scenario, the observed jet (as usually defined) is originated
     from the precession of the single mini-jet which ejects magnetized plasma
     and superlumnal plasmoids in a long period.\\
      Our precessing nozzle scenario can be understood in the framework of
     available relativistic magnetohydrodynamic theories for the 
     formation/collimation/acceleration  of relativistic jets in blazars 
     (cf. a detailed discussion in Qian et al. \cite{Qi17}). 
     The structure of the jets observed  in the radio galaxy M87 (parabolic
     hollow jet nozzle structure; Nakamura \& Asada \cite{Na13}) and 
     in OJ287 (fork-like jet structure; Tateyama \cite{Ta13}) may be 
     regarded as observational evidence.\\
     Theoretically, magnetic nozzles can be probably formed in disk-driven 
     jets in the magnetospheres  of rotating  black-hole/accretion-disk 
     systems. These magnetic nozzles may locate near the classical fast 
     magnetosonic point where the magnetohydrodynamic  flow remains  Poynting 
     flux dominated. In some self-similar  axisymmetric MHD flow models,
     beyond the classic fast magnetosonic point, the jet will be accelerated
     untill approaching the modified fast magnetosonic point (Blandford \&
     Znajek \cite{Bl77}; Li et al. \cite{Li92}; Vlahakis \& K\"onigl 
     \cite{Vl04}). This extended acceleration is due to the dominance of 
     Poynting flux at the classic magnetosonic point, thus having ample 
     electromagnetic energy to be transformed plasma kinetic energy (Komissarov
    et al. \cite{Kom07}, Komissarov \cite{Kom09}, Millas et al. \cite{Mi14}).\\
   . In the case of relativistic jets, the classical
   fast magnetosonic point is located in the force-free region of the 
    magnetosphere where the magnetic energy dominates the plasma kinetic
    energy. The magnetic field lines anchored
    into the innermost disk and the magnetic nozzle would rotate rigidly
    with the disk (MacDonald \& Throne \cite{Mac82}). \\
    Additionally, in this extended acceleration region beyond the magnetic
     nozzle (or beyond the classical fast magnetosonic point), 
    the inertia of the plasma becomes strong and the 
    electromagnetic fields are neither degenerate nor force free, and the
    plasma would flow  along its own streamlines, not following the 
   local field lines and forming its own trajectory pattern.
   This can explain  why the superluminal knots observed in B1308+326
    move along the common precessing track.\footnote{Also in other QSOs and
    blazars, (e.g., in PG 1302-102, NRAO 150, OJ287, 3C345, 3C454.3 
    and 3C279.} \\
   Most recently, the black-hole/accretion-disk system in the giant
   radio galaxy M87 (having a hole-mass of  
    $\sim$6.5$\times{10^9}$$M_{\odot}$) has been observed by using the 
   world mm-VLBI-network (Lu et al. \cite{Lu23}).
    Both the mm-jet emanating from its centeral 
   hole and the associated circumdisk structure  have been mapped.
   The morphological strcuture and kinematic properties are well
   consistent with the whole picture expected by relativistic MHD 
   theories for jet formation in black-hole/accretion disk systems.\\


 \end{document}